\documentclass[article]{elsarticle}
\usepackage[margin=2.5cm]{geometry}
\usepackage{lineno,hyperref}
\modulolinenumbers[5]
\pdfoutput=1 % if your are submitting a pdflatex (i.e. if you have
\usepackage[utf8]{inputenc}% Para ecentos y demas
\journal{.}
\usepackage{amssymb}
\usepackage{amsthm}
\usepackage{amsxtra}
\usepackage{stmaryrd}
\usepackage{mathrsfs}
\usepackage[scr,scaled=1.1]{rsfso}
\numberwithin{equation}{section}
\usepackage{stmaryrd}
\usepackage{eucal}
\usepackage{shuffle}
\newcommand{\Ascr}{\mathscr{A}}
\newcommand{\Dscr}{\mathscr{D}}
\newcommand{\Mscr}{\mathscr{M}}
\newcommand{\Wscr}{\mathscr{W}}
\newcommand{\OWscr}{\mathscr{OW}}

\newcommand{\Acal}{\mathcal{A}}
\newcommand{\Gcal}{\mathcal{G}}

\newcommand{\ue}{\operatorname{e}}
\newcommand{\ui}{\mathrm{i}\mkern1mu}

\def\cs{\mathsf{cs}}
\def\cd{\mathsf{cd}}

\def\half{{1\over 2}}

\def\Z{{\mathchoice {\hbox{$\sf\textstyle Z\kern-0.4em Z$}}
{\hbox{$\sf\textstyle Z\kern-0.4em Z$}}
{\hbox{$\sf\scriptstyle Z\kern-0.3em Z$}}
{\hbox{$\sf\scriptscriptstyle Z\kern-0.2em Z$}}}}

\def\square{\kern1pt\vbox{\hrule height 1.2pt\hbox{\vrule width 1.2pt
   \hskip 3pt\vbox{\vskip 6pt}\hskip 3pt\vrule width 0.6pt}
   \hrule height 0.6pt}\kern1pt}
      \def\boxop{{\raise-.25ex\hbox{\square}}}
\def\tr{{\rm tr}\,}

\def\partder#1#2{{\partial #1\over\partial #2}}

\newcommand{\be}{\begin{equation}}
\newcommand{\ee}{\end{equation}\noindent}
\newcommand{\bear}{\begin{eqnarray}}
\newcommand{\ear}{\end{eqnarray}\noindent}
\newcommand{\benn}{\begin{enumerate}}
\newcommand{\enn}{\end{enumerate}}

\def\slash#1{#1\!\!\!\raise.15ex\hbox {/}}
\newcommand{\slD}{\,\raise.15ex\hbox{$/$}\kern-.27em\hbox{$\!\!\!D$}}
\newcommand{\slpartial}{\raise.15ex\hbox{$/$}\kern-.57em\hbox{$\partial$}}
\def\epseps#1#2{\varepsilon_{#1}\cdot \varepsilon_{#2}}
\def\epsk#1#2{\varepsilon_{#1}\cdot k_{#2}}

\def\Gdb{{{\dot {\bar G}}}}
\def\Gddb{{{\ddot {\bar G}}}}
\def\Gdb{{{\dot {\bar G}}}}

\def\Gddb{{{\ddot {\bar G}}}}

\def\Gbd{\dot G}

\def\4piTD{{(4\pi T)}^{-{D\over 2}}}
\def\4piT4{{(4\pi T)}^{-2}}

\def\Tintm4{{\dps\int_{0}^{\infty}}{dT\over T}\,e^{-m^2T}
    {(4\pi T)}^{-2}}
\def\Tintm{{\dps\int_{0}^{\infty}}{dT\over T}\,e^{-m^2T}}

\newcommand{\slG}{{{\dot G}\!\!\!\! \raise.15ex\hbox {/}}}
\newcommand{\Gd}{{\dot G}}

\def\GBd12{{\dot G}_{12}}

\newcommand{\no}{\noindent}

\def\dps{\displaystyle}

\newcommand{\bea}{\begin{eqnarray}}  
\newcommand{\eea}{\end{eqnarray}}

\def\sqr#1#2{{\vcenter{\vbox{\hrule height.#2pt  
     \hbox{\vrule width.#2pt height#1pt \kern#1pt  
           \vrule width.#2pt}  
       \hrule height.#2pt}}}}  
\def\square{\mathchoice\sqr66\sqr66\sqr{2.1}3\sqr{1.5}3}  
\def\appendix{\par\clearpage
  \setcounter{section}{0}
  \setcounter{subsection}{0}
  \def\@sectname{Appendix~}
  \def\theequation{\thesection\arabic{equation}}
  \def\thesection{\Alph{section}}}
 
\def\thefigures#1{\par\clearpage\section*{Figures\@mkboth
  {FIGURES}{FIGURES}}\list
  {Fig.~\arabic{enumi}.}{\labelwidth\parindent\advance
\labelwidth -\labelsep
      \leftmargin\parindent\usecounter{enumi}}}

\def\thetables#1{\par\clearpage\section*{Tables\@mkboth
  {TABLES}{TABLES}}\list
  {Table~\Roman{enumi}.}{\labelwidth-\labelsep
      \leftmargin0pt\usecounter{enumi}}}

\def\@sect#1#2#3#4#5#6[#7]#8{\ifnum #2>\c@secnumdepth
     \def\@svsec{}\else
     \refstepcounter{#1}\edef\@svsec{\@sectname\csname the#1\endcsname
.\hskip 1em }\fi
     \@tempskipa #5\relax
      \ifdim \@tempskipa>\z@
        \begingroup #6\relax
          \@hangfrom{\hskip #3\relax\@svsec}{\interlinepenalty \@M #8\par}
        \endgroup
       \csname #1mark\endcsname{#7}\addcontentsline
         {toc}{#1}{\ifnum #2>\c@secnumdepth \else
                      \protect\numberline{\csname the#1\endcsname}\fi
                    #7}\else
        \def\@svse=chd{#6\hskip #3\@svsec #8\csname #1mark\endcsname
                      {#7}\addcontentsline
                           {toc}{#1}{\ifnum #2>\c@secnumdepth \else
                             \protect\numberline{\csname the#1\endcsname}\fi
                       #7}}\fi
     \@xsect{#5}}
 
\def\@sectname{}
\def\partder#1#2{{\partial #1\over\partial #2}}

\newcommand{\nc}{\newcommand}
\nc{\spa}[3]{\left\langle#1\,#3\right\rangle}
\nc{\spb}[3]{\left[#1\,#3\right]}
\nc{\ksl}{\not{\hbox{\kern-2.3pt $k$}}}
\nc{\hf}{\textstyle{1\over2}}
\nc{\pol}{\varepsilon}
\nc{\tq}{{\tilde q}}
\nc{\esl}{\not{\hbox{\kern-2.3pt $\pol$}}}
\renewcommand{\theequation}{\arabic{section}.\arabic{equation}}

\def\GBd12{{\dot G}_{12}}

\def\beqn*{\begin{eqnarray*}}
\def\eqn*{\end{eqnarray*}}

\def\square{\kern1pt\vbox{\hrule height 1.2pt\hbox{\vrule width 1.2pt
   \hskip 3pt\vbox{\vskip 6pt}\hskip 3pt\vrule width 0.6pt}
   \hrule height 0.6pt}\kern1pt}

\def\slash#1{#1\!\!\!\raise.15ex\hbox {/}}

\def\dps{\displaystyle}

\def\half{{1\over 2}}

\def\epseps#1#2{\varepsilon_{#1}\cdot \varepsilon_{#2}}
\def\epsk#1#2{\varepsilon_{#1}\cdot k_{#2}}

\def\4piTD{{(4\pi T)}^{-{D\over 2}}}
\def\4piT4{{(4\pi T)}^{-2}}

\def\Tintm4{{\dps\int_{0}^{\infty}}{dT\over T}\,e^{-m^2T}
    {(4\pi T)}^{-2}}
\def\Tintm{{\dps\int_{0}^{\infty}}{dT\over T}\,e^{-m^2T}}

\def\tr{{\rm tr}\,}

\def\oplusotimes{{{\lower 15pt\hbox{$\scriptscriptstyle \oplus$}}\atop{\otimes}}}
\def\perppar{{{\lower 15pt\hbox{$\scriptscriptstyle \perp$}}\atop{\parallel}}}
\def\oopp{{{\lower 15pt\hbox{$\scriptscriptstyle \oplus$}}\atop{\otimes}}\!{{\lower 15pt\hbox{$\scriptscriptstyle \perp$}}\atop{\parallel}}}
\def\bbbz{{\mathchoice {\hbox{$\sf\textstyle Z\kern-0.4em Z$}}
{\hbox{$\sf\textstyle Z\kern-0.4em Z$}}
{\hbox{$\sf\scriptstyle Z\kern-0.3em Z$}}
{\hbox{$\sf\scriptscriptstyle Z\kern-0.2em Z$}}}}

\begin{document}

\begin{frontmatter}

\title{Manifest colour-kinematics duality and double-copy in the string-based formalism}
\author[a]{Naser Ahmadiniaz}
\ead{n.ahmadiniaz@hzdr.de}
\address[a]{Helmholtz-Zentrum Dresden-Rossendorf, Bautzner Landstra\ss e 400, 01328 Dresden, Germany}
\author[b,c]{Filippo Maria Balli}
\ead{filippo.balli@unimore.it}
\address[b]{Dipartimento di Scienze Fisiche, Informatiche e Matematiche, Universit\`{a} degli Studi di Modena e Reggio Emilia, Via Campi 213/A, I-41125 Modena, Italy\\
and INFN, Sezione di Bologna, Via Irnerio 46, I-40126 Bologna, Italy}
\address[c]{Department of Physics and Astronomy, Uppsala University, 75108 Uppsala, Sweden}
\author[b]{Olindo Corradini}
\ead{olindo.corradini@unimore.it}
%\address[b]{Dipartimento di Scienze Fisiche, Informatiche e Matematiche, Universit\`{a} degli Studi di Modena e Reggio Emilia, Via Campi 213/A, I-41125 Modena, Italy\\
%and INFN, Sezione di Bologna, Via Irnerio 46, I-40126 Bologna, Italy}
% more complex case: 4 authors, 3 institutions, 2 footnotes
\author[d]{Cristhiam Lopez-Arcos\corref{correspondingauthor}}
\cortext[correspondingauthor]{Corresponding author}
\ead{cmlopeza@unal.edu.co}
\author[d]{Alexander Quintero V\'{e}lez}
\ead{aquinte2@unal.edu.co}
\address[d]{Escuela de Matem\'{a}ticas, Universidad Nacional de Colombia Sede Medell\'{i}n, Carrera 65 $\#$ 59A--110, Medell\'{i}n, Colombia}
\author[e]{Christian Schubert}
\ead{schubert@ifm.umich.mx}
\address[e]{Instituto de F\'{i}sica y Matem\'{a}ticas Universidad Michoacana de San Nicol\'{a}s de Hidalgo
Edificio C-3, Apdo. Postal 2-82 C.P. 58040, Morelia, Michoac\'{a}n, M\'{e}xico}

\begin{abstract}

The relation for the gravity polarisation tensor as the tensor product of two gluon polarisation vectors has been well-known for a long time, but a version of this relation for multi-particle fields is presently still not known. Here we show that in order for this to happen we first have to ensure that the multi-particle polarisations satisfy colour-kinematics duality. In previous work it has been show that this arises naturally from the Bern-Kosower formalism for one-loop gluon amplitudes, and here we show that the tensor product for multi-particle fields arise naturally in the Bern-Dunbar-Shimada formalism for one-loop gravity amplitudes. This allows us to formulate a new prescription for double-copy gravity Berends-Giele currents, and to obtain both the colour-dressed Yang-Mills Berends-Giele currents in the Bern-Carrasco-Johansson gauge and the gravitational Berends-Giele currents explicitly. An attractive feature of our formalism is that it never becomes necessary to determine gauge transformation terms. Our double-copy prescription can also be applied to other cases, and to make this point we derive the double-copy perturbiners for $\alpha'$-deformed gravity and the bi-adjoint scalar model.

\end{abstract}

%\begin{keyword}
%\texttt{elsarticle.cls}\sep \LaTeX\sep Elsevier \sep template
%\MSC[2010] 00-01\sep  99-00
%\end{keyword}

\end{frontmatter}
\begin{minipage}{\textwidth}
\vspace{-280mm}{\flushright UUITP-47/21 \\}
\end{minipage} 
%\addvspace{-400mm}{\flushright UUITP-47/21 \\}

%\vspace*{-100mm}{\flushright UUITP-47/21 \\}

%\linenumbers

\tableofcontents

\section{Introduction}\label{sec:intro}
The perturbative expansion of Yang-Mills theory has remarkable aspects and properties which are extremely difficult to be seen and extracted from the Feynman rules. Parke and Taylor \cite{Parke:1986gb} made a conjucture about the tree-level scattering amplitudes that are maximally helicity violating (MHV) to be expressed in terms of simple holomorphic functions. Their observation was based on a computation for the first few cases but later was proved by Berends and Giele \cite{Berends:1987me} for the general case. 
Recent years have seen a tremendous development in the area of the on-shell matrix elements calculation in quantum field theory, especially for gauge theory and gravity, such as unitary-based methods \cite{Bern:1994zx,Bern:2011qt}, twistors~\cite{Witten:2003nn}, Britto-Cachazo-Feng-Witten (BCFW) recursion
\cite{Britto:2004ap,Britto:2005fq} and Grassmannians \cite{Arkani-Hamed:2009ljj,Mason:2009qx}, see \cite{Elvang:2013cua,Dixon:2013uaa} for recent reviews.   

In an independent development, the perturbiner expansion method  was introduced by Rosly and Selivanov~\cite{Rosly:1996vr, Rosly:1997ap, Selivanov:1997aq, Selivanov:1997ts} as an efficient method to obtain the tree-level scattering amplitudes for a generic massless quantum field theory and, in the Yang-Mills case, can as well be used as a tool to compute multi-particle trees with one particle off-shell, i.e. the so-called Berends-Giele currents~\cite{Berends:1987me}. In fact, it turns out that these currents can be efficiently packed if a particular gauge is chosen, namely the Bern-Carrasco-Johansson (BCJ) gauge~\cite{Bern:2008qj}, which displays the very helpful feature known as ``colour-kinematics duality'', where colour-ordered amplitudes obey the same relations as their associated colour factors. In the Berends-Giele currents, colour-kinematics duality was made explicit by identifying multi-particle polarisations which satisfy the so-called generalized Jacobi identities (GJI)~\cite{Mafra:2014oia}.

In order to extend such constructions to gravity, a possible general strategy which has been widely used is to rely upon  perturbative gauge-gravity duality, which is a consequence of open-closed duality of string theory and at the perturbative string level gives rise to the so-called Kawai-Lewellen-Tye (KLT) relations between open string amplitudes and closed string amplitudes~\cite{Kawai:1985xq}.  In the particle limit of string theory ($\alpha'\rightarrow0$), it leads to relations between tree-level graviton amplitudes and tree-level gluon amplitudes in Yang-Mills theories, which are often summarized as ``${\rm gravity}=({\rm gauge~theory})^2$". Such duality holds even though the structures of the non-abelian Yang-Mills and the Einstein-Hilbert lagrangians are rather different: the former contains only up to four-point interactions while the latter contains infinitely many vertices. Therefore the validity of the above duality in the field theory limit has been a major puzzle for many years. Finally, however powerful, these rules are limited to relations between scattering amplitudes, i.e. they are on-shell expressions.

The original derivation of the KLT relations was done in string theory \cite{Kawai:1985xq}. Later on they were applied
 to $n$-point amplitudes in field theories in \cite{Bern:1998sv,Bjerrum-Bohr:2010pnr}, and more recently they were revisited in a more algebro-topological framework in \cite{Mizera:2017cqs}. The connection between gravity and gauge theory starts already at three points. The three-point amplitudes vanish for real on-shell momenta, 
 but this can be avoided by going to complex momenta which leads to the following compact result
\bear
\mathscr{M}_{3}(1,2,3)=\mathscr{A}_3(1,2,3)\tilde{\mathscr{A}}_3(1,2,3)
\ear
where $\mathscr{M}_3$ and $\mathscr{A}_3 (\tilde{\mathscr{A}}_3)$ are the three-point gravity and gauge theory amplitudes accordingly \cite{Sondergaard:2011iv}. For the four- and five-point amplitudes 
the relations look slightly different but still very simple 
\bear
\mathscr{M}_4(1,2,3,4)&=&-s_{12}\mathscr{A}_4(1,2,3,4)\tilde{\mathscr{A}}_{4}(1,2,4,3)\nonumber\\
\mathscr{M}_{5}(1,2,3,4,5)&=&s_{12}s_{34}\mathscr{A}_5(1,2,3,4,5)\tilde{\mathscr{A}}_5(2,1,4,3,5)+s_{13}s_{24}\mathscr{A}_5(1,3,2,4,5)\tilde{\mathscr{A}}_5(3,1,4,2,5)
\ear  
where $s_{ij}=(k_i+k_j)^2=2k_i\cdot k_j$.

Bern, Carrasco and Johansson~\cite{Bern:2008qj} discovered a direct way of constructing gravity amplitudes from gauge theory amplitudes after organizing the latter in a specific manner, so that the amplitude numerator respects a certain colour-kinematics duality, which involves colour factors and kinematic numerators; then by a suitable replacement which is called the ``double-copy construction'', they obtained tree-level gravity amplitudes. 
Specifically, at tree-level, colour dressed scattering amplitudes in Yang-Mills theories can be written in the following form
\begin{equation}\label{eq:CK-amp}
\mathscr{A}_n \sim \sum_{i} \frac{c_in_i}{D_i}
\end{equation}
where the $c_i$ are the colour factors or numerators consisting of contractions of the gauge Lie algebra structure constants, the $n_i$ are the kinematic numerators which are sums of Lorentz-invariant contractions of external momenta and polarisations, and the $D_i$ are the propagators which can be written in term of Mandelstam variables. One particularity is that the sum is only over cubic or trivalent graphs with the quartic vertices absorbed by the cubic ones. Colour-kinematics duality means that the relations satisfied by the colour factors due to Jacobi identities are mirrored by their respective kinematic numerators, i.e. 
\bear
c_i+c_j+c_k=0\ \rightarrow\ n_i+n_j+n_k=0.
\ear

The great advantage of having the amplitude represented like \eqref{eq:CK-amp} is that the calculation of the associated gravity amplitude is automatic, as soon as the numerators satisfy colour-kinematics. The gravity amplitudes are obtained in terms of the gauge theory information simply by replacing the colour factors by another copy of the kinematic numerators and summing over the same cubic diagrams. The $n$-point gravity amplitude is given by 
\bear
\mathscr{M}_n\sim\sum_{i} \frac{n_i\tilde{n}_i}{D_i}.
\ear
Interestingly, the double-copy construction has also been used in more phenomenological studies, such as in gravitational wave physics (see for expample Refs. \cite{Goldberger:2017frp,Shen:2018ebu,Cheung:2018wkq}) which since the direct detection of gravitational waves in the LIGO and VIRGO experiments \cite{LIGOScientific:2016aoc,LIGOScientific:2017vwq} has become a very active field of research. 
Very recently, a classical double-copy relation was developed in the context of a worldline QFT description
of the classical gravitational scattering of massive bodies \cite{Shi:2021qsb}.

In general, in the literature there are several proofs for the BCJ amplitude relations from different approaches such as the string-theory monodromy relations \cite{Bjerrum-Bohr:2009ulz,Stieberger:2009hq,Bjerrum-Bohr:2010mia} and the BCFW recursion relations \cite{Britto:2004ap,Britto:2005fq}. In the field theory limit, in contrast to the KLT relations, the double-copy construction has been conjectured to hold for integrands of loop-level amplitudes \cite{Bern:2019prr} which has been tested in various publications \cite{Bern:2010ue,Carrasco:2011mn,Carrasco:2011hw,Bern:2011rj,Boucher-Veronneau:2011rlc,Naculich:2011my,Bern:2012uf}. There is a lot of evidence in supporting the conjucture that colour-kinematic dual representations exist for a variety of gauge theory amplitudes. This fact has been shown explicitly for tree-level amplitudes with up to eight external legs \cite{Bern:2019prr}, 
and for integrands at several loop-levels in maximally Supersymmetric-Yang-Mills theory \cite{Bern:2008qj,Bern:2010ue,Carrasco:2011mn,Bern:2012uf}. Less supersymmetric theories are also expected to respect such duality at both tree and loop orders, see \cite{Bern:2019prr} for a two-loop QCD example. In Ref.~\cite{Bargheer:2012gv} the duality between colour and kinematics has been generalized to three-algebras in three-dimensional supersymmetric Chern-Simons theories. Double-copy also found applicability in anyonic models \cite{Burger:2021wss}.

At loop level, starting from string amplitudes, Bern and Kosower unveiled a very powerful master formula for the one-loop gluon scattering amplitudes~\cite{Bern1991-1669,bern1991color,bern1992computation}, later re-derived by Strassler directly from the particle approach, i.e. from the worldline formalism~\cite{Strassler:1992zr, Strassler:1992nc}, which does not require on-shell conditions nor masslessness. Bern, Dunbar and Shimada then extended the Bern-Kosower string-based rules to gravity~\cite{Bern:1993wt, Dunbar:1994bn}. Their approach made essential use of the aforementioned open-closed string duality, treating graviton amplitudes as a {\it double-copy} of gluon amplitudes. Such approach is reviewed below, in Section~\ref{sec:BK-BDS}. 

A common difficulty in both scenarios, i.e. gluon and gravity amplitudes, is the explicit realization of Ward identities, which in the off-shell case relate $n$-particle amplitudes to $(n-1)$-particle amplitudes. In the on-shell case they are encoded in the transversality condition which involves both one-particle irreducible and reducible diagrams, which arise when trees are sewn to the loop. Hence, it is of fundamental importance to have a method which allows one to efficiently compute these trees. As already mentioned in the gluon case such multi-particle trees, with
one particle off-shell, are dubbed Berends-Giele currents~\cite{Berends:1987me}.

Recently, some of the present authors investigated the construction of Berends-Giele currents for gluons, inspired by the Bern-Kosower replacement rules~\cite{Ahmadiniaz:2021fey}. The most relevant aspect of the Bern-Kosower rules to such purpose is a procedure known as ``pinching procedure'' that allows one to construct the reducible parts of the amplitudes from the irreducible parts of the one-loop amplitudes at the level of the Feynman-Schwinger integrands. First one employs suitable integrations by parts which effectively remove quartic vertices and yield integrands which are expressed in terms of ``Lorentz cycles'', i.e. traces of products of the linearized parts of gluon field strenghts in momentum space. Then, one suitably removes linearly appearing Green's functions, which correspond to the external lines to be pinched. In \cite{Ahmadiniaz:2021fey} this pinching procedure was implemented with the introduction of a differential operator, so-called ``pinch operator'', and 
also the parts of the integrand identified that have to be pinched in order to extract the multi-particle polarisations that will conform the Berends-Giele currents. Such polarisations come naturally in the BCJ gauge. The results of the above paper are summarized in Section~\ref{sec:CK}.
 
In the present manuscript, in Section~\ref{sec:DC}, we extend such a construction to gravity amplitudes, by considering the Bern-Dunbar-Shimada formalism and the double-copy procedure at the level of the Berends-Giele currents. In other words, we find the multi-particle polarisation expansion for gravity by means of the ``double pinch operator'' which is a double-copy of the one used in the Yang-Mills case. These multi-particle polarisation currents are thus used to construct the perturbiner expansion. Unlike other recent approaches~\cite{Mizera:2018jbh, Cho:2021nim}, in our method the off-shell double-copy polarisation currents emerge directly from the Bern-Dunbar-Shimada formalism, rather than from KLT relations.

Our new prescription for the double-copy Berends-Giele currents can be applied to other models. In Section \ref{sec:examples} we present the perturbiners for the cases of the $\alpha'$-deformed gravity and the bi-adjoint scalar model. Finally, in Section \ref{sec:concl}, we conclude.

\paragraph{Notation}
Latin indices $a,b$, etc. from the beginning of the alphabet run over the $N^{2}-1$ generators of $\mathfrak{su}(N)$. We let these generators be denoted by $T_{a}$, with structure constants $f_{ab}^{\phantom{ab}c}$ satisfying 
$$
[T_{a},T_{b}] = \ui \sqrt{2} f_{ab}^{\phantom{ab}c} T_{c}.
$$ 
For reasons of formal convenience, we set $\tilde{f}_{ab}^{\phantom{ab}c} =  \ui \sqrt{2} f_{ab}^{\phantom{ab}c}$. 

By a word we mean a finite string $P=i_1 i_2 \cdots i_n$ of positive integers $i_1,i_2,\dots,i_n \geq 1$. The word consisting of no symbols is called the empty word, written $\varnothing$. Given a word $P=i_1 i_2 \cdots i_n$, we denote by $\lvert P \rvert$ its length $n$. Also, the multi-particle momentum for such word and its associated Mandelstam invariant are given by $k_{P} = k_{i_1} + k_{i_2} + \cdots +  k_{i_n}$ and $s_P = k_P^2$.  

We shall implicitly work with the free Lie algebra generated by all words with letters in $1,2, \dots, n$. There one can consider the left-to-right bracketing $\ell$, which is defined recursively by
\begin{align*}
\ell(\varnothing) &= 0, \\
\ell(i) &= i, \\
\ell(i_1i_2 \cdots i_n) &= \ell(i_1i_2 \cdots i_{n-1})i_n -  i_n \ell(i_1i_2 \cdots i_{n-1}).   
\end{align*}
Using this notation, for objects labelled by words, the generalised Jacobi identities of order $k$ can be characterised as
$$
U_{P \ell(Q)} + U_{Q \ell(P)} = 0
$$
for every pair of non-empty words $P$ and $Q$ such that $\lvert P \rvert + \lvert Q \rvert = k$. For example, 
\begin{align}\label{eq:GJI}
\begin{split}
U_{12} + U_{21} &= 0, \\
U_{123} + U_{312} + U_{231} &= 0, \\
U_{1234} + U_{2143} + U_{3412} + U_{4321} &=0.
\end{split}
\end{align}
If $U_P$ satisfies the generalised Jacobi identities we will use the notation $U_{\ell(P)}$ instead of $U_P$. In particular, this implies that $U_{[P,Q]} = U_{P \ell(Q)}$.

%%%%%%%%%%%%%%%%%%%%%%%%%%%%%%%%%%%%%%%%%%%%%%%%%%%%%%%%%%%%%%%%%%%%%%%%%%%%%%%%%%%%
\section{The string-based rules in field theory}\label{sec:BK-BDS}

Bern and Kosower \cite{bern1991color,bern1992computation}
 were the first to systematically investigate the usefulness of the fact that
many amplitudes in field theory can be represented as the infinite string tension limit of 
appropriately chosen string amplitudes. In this way they obtained rules for the construction of
Feynman-Schwinger type parameter integral representations 
of the one-loop on-shell $n$-gluon amplitudes with a scalar, spinor or gluon loop. 
This work was then generalized to the one-loop $n$-graviton amplitudes by
Bern, Dunbar and Shimada \cite{Bern:1993wt} 
with some details filled in later by Dunbar and Norridge \cite{Dunbar:1994bn}.

\subsection{The one-loop $n$-gluon amplitudes}

The central object in the Bern-Kosower formalism is the colour-ordered one-loop $n$-gluon correlator with a massless scalar loop. This amplitude in general has a one-particle irreducible and a one-particle reducible contribution. The irreducible one is encoded in the following master formula,
%
%\begin{widetext}
\begin{eqnarray}
\Gamma(k_1,\varepsilon_1;\ldots;k_n,\varepsilon_n)
&=&
{(-ig)}^n
\tr (T^{a_1}\cdots T^{a_n})
{\int_{0}^{\infty}}dT
{(4\pi T)}^{-\frac{D}{2}}
\int_0^T d\tau_1 \int_0^{\tau_1}d\tau_2 \cdots \int_0^{\tau_{n-2}} d\tau_{n-1}
\nonumber\\ &&\times 
\exp \biggl\lbrace \sum_{i,j=1}^n 
\Bigl(  \half G_{ij} k_i\cdot k_j
-i\dot G_{ij}\varepsilon_i\cdot k_j
+\half\ddot G_{ij}\varepsilon_i\cdot\varepsilon_j
\Bigr)\biggr\rbrace
\Bigl\vert_{\varepsilon_1\ldots \varepsilon_n}
\label{gluonmaster}
\end{eqnarray}
%\end{widetext}
\no
Here 
\bear
G_{ij}\equiv G(\tau_i,\tau_j) = \vert \tau_i -\tau_j\vert -\frac{(\tau_i-\tau_j)^2}{T}
\ear
so that 
\bear
\Gd_{ij}\equiv \partder{}{\tau_i} G_{ij} = {\rm sign}(\tau_i-\tau_j) - 2\frac{ (\tau_i-\tau_j)}{T}
\ear
(the explicit form of $\ddot G_{ij}\equiv \frac{\partial^2}{\partial\tau_i^2}G_{ij}$ is not needed). The notation
$\bigl\vert_{\varepsilon_1\ldots \varepsilon_n}$ means that the exponential
has to be expanded keeping only the terms
linear in each of the polarisation vectors $\varepsilon_1,\ldots,\varepsilon_n$. 
The resulting integrand is of the form 
\bear \exp\biggl\lbrace 
\cdot
\biggr\rbrace 
\Bigl\vert_{\varepsilon_1\varepsilon_2\ldots \varepsilon_n}
 \equiv
 {(-i)}^n P_n(\dot G_{ij},\ddot G_{ij})
 {\rm e}^{\half \sum_{i,j=1}^n G_{ij}k_i\cdot k_j }
 \label{defPN}
 \ear\no
 with certain polynomials $P_n$.

\begin{figure}[h]
\centering 
\includegraphics[scale=0.85]{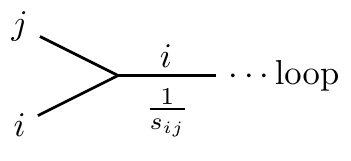}
\caption{\label{fig-pinch} Pinching of a vertex according to the Bern-Kosower rules.}
\end{figure}

The reducible contributions can be included by the following ``pinching procedure'':
(i) Remove the second derivatives $\ddot G_{ij}$ contained in $P_n$ through suitable partial integrations. 
This step will lead to the replacement $P_n(\dot G_{ij},\ddot G_{ij})\ \to Q_n(\dot G_{ij})$.
(ii) Draw all possible $\phi^3$ one-loop diagrams $D_i$ with $n$ legs, labelled $1,\ldots,n$ and following the
ordering of the colour trace. 
(iii) The pinching rule amounts to the replacement
\bear
\dot G_{ij} \longrightarrow \frac{2}{s_{ij}}=\frac{2}{(k_i+k_j)^2}
\label{pinchij}
\ear
removing the vertex and transferring the label $i$ to the ingoing leg (see Fig. \ref{fig-pinch}).  

\noindent
The $\tau_j$ - integration is omitted and the index $j$ replaced by $i$ in the remaining $G_{kl}$ and $\dot G_{kl}$. (iv) The previous replacement can only occur on a vertex with labels $i<j$, iff $Q_n$ contains $\dot G_{ij}$ linearly. Moreover, a diagram will contribute iff each vertex except the ones attached directly to the loop corresponds to a possible pinch. The pinching procedure starts with the outermost vertices and recursively removes the trees attached to the loop.  

As a further benefit of the integration-by-parts procedure, at this stage the contributions of the spinor and gluon 
loop to the $N$-gluon amplitudes can be constructed at the integrand level using a set of ``loop replacement rules''
\cite{bern1991color,bern1992computation}.

The pinching procedure was streamlined in \cite{Ahmadiniaz:2021fey} by the introduction of a \emph{pinch operator}, that we will present here again in the next section, and the implementation of the multi-particle techniques that helped to have a better understanding of the structure of the trees attached to the loop.

\subsection{The one-loop $n$-graviton amplitudes}

The gluon master formula \eqref{gluonmaster} was generalized in \cite{Bern:1993wt,Dunbar:1994bn} to a master formula for the
irreducible one-loop $n$-graviton amplitudes with a massless scalar loop:
\bear
\Gamma[k_1,h_1;\cdots ;k_n,h_n]
&=&
%(2\pi)^D\delta(\sum k_i)
-(-{\kappa\over 4})^n
\int_0^{\infty}{dT\over T}(4\pi T)^{-{D\over 2}}
\int_0^Td\tau_1\cdots \int_0^Td\tau_n\nonumber\\
&&\times
\,\exp\Biggl\lbrace
\sum_{i,j=1}^n
\biggl\lbrack
\half G_{ij}k_i\cdot k_j -i (\dot G_{ij}\varepsilon_i+\dot {\bar G}_{ij}\bar\varepsilon_i)\cdot k_j
+\half \ddot G_{ij} \varepsilon_i\cdot \varepsilon_j
\nonumber\\
&& +\half \ddot {\bar G}_{ij}\bar\varepsilon_i \cdot \bar\varepsilon_j
+\half H_{ij}(\varepsilon_i\cdot \bar\varepsilon_j + \varepsilon_j\cdot \bar\varepsilon_i)
\biggr\rbrack\Biggr\rbrace
\Bigl\vert_{\varepsilon_1\ldots \varepsilon_n\bar\varepsilon_1\ldots\bar\varepsilon_n}
\label{mastergrav}
\ear
Here we have used that on-shell the graviton polarisations can be chosen so as to factorize,
$h_i^{\mu\nu}=\varepsilon_i^\mu\bar\varepsilon_i^\nu$. In the absence of the terms with $H_{ij}$ 
this would, after the expansion of the exponent, lead to a prefactor polynomial that simply factorizes 
into two copies of the one of the gluonic case in \eqref{defPN},
\bear \exp\biggl\lbrace 
\cdot
\biggr\rbrace 
\Bigl\vert_{\varepsilon_1\ldots \varepsilon_n\bar\varepsilon_1\ldots\bar\varepsilon_n}
 =
 P_n(\dot {\bar G}_{ij},\ddot {\bar G}_{ij})
 P_n(\dot G_{ij},\ddot G_{ij})
  {\rm e}^{\half \sum_{i,j=1}^n G_{ij}k_i\cdot k_j }
 \label{expandgrav}
 \ear\no
 At the string level, this comes from the factorisation of the closed string modes into left-movers and right-movers. 
The additional terms involving $H_{ij}$ stem from the fact that the left- and right-movers 
are coupled through the zero mode of the string. In order to study the structure of the now gravitational trees attached to the loop, \eqref{expandgrav} is all we need, but for the calculation of the whole one-loop irreducible part we cannot neglect the contributions that come from these terms. 

Differently from the gluon case, it is now generally not possible to remove all of the $\dot G_{ij}, \dot {\bar G}_{ij}$ 
using partial integrations in the variables $\tau_i$ alone. Instead, one has to return to the string level and appeal
to the fact that, before taking the infinite string tension limit, the left-and right movers depended on independent variables
$\tau_i$ and $\bar \tau_i$. This allows one to write $\dot G_{ij} = \partder{}{\tau_i}G_{ij}$, 
$\dot {\bar G}_{ij} = \partder{}{\bar\tau_i}{\bar G}_{ij}$ and to treat
$\dot G_{ij}, \ddot G_{ij}$ as independent of $\dot {\bar G}_{ij}, \ddot {\bar G}_{ij}$
in the partial integration procedure. Additionally, the following rules must be used for derivatives
hitting the universal exponent, 
\bear
\frac{\partial}{\partial{\bar \tau_k}}
\dot G_{ij}  &=& \half(\delta_{ki}H_{ij}-\delta_{kj}H_{ij}) 
\\
{\partial \over \partial{\tau_k} } \Gdb_{ij} 
&=&
\half(\delta_{ki}H_{ij}-\delta_{kj}H_{ij})
 \\
{\partial\over \partial{\bar \tau_k} } \ddot G_{ij} &=& 0
\\
{ \partial\over \partial{ \tau_k} }\Gddb_{ij} &=& 0
\label{crossterms}
\ear
The $H_{ij}$ are to be treated as constants in the integration-by-parts. 

After the removal of the $\ddot G_{ij}, \ddot {\bar G}_{ij}$, the inclusion of the reducible contributions can be
achieved by a pinching procedure that parallels the one for the gluon case above, except that the condition
for the pinching of a vertex with labels $i<j$ now is that the integrand should contain both $\dot G_{ij}$ and 
$\dot {\bar G}_{ij}$ linearly, and that the replacement \eqref{pinchij} has to be modified to
\bear
\Gbd_{ij} \Gdb_{ij}
\rightarrow  
\frac{4}{s_{ij}}
% i \rightarrow j \hskip 2 cm \hbox{in remaining factors} }
\ear
After the recursive removal of all trees attached to the loop one has at hand a parameter integral representation
for the full on-shell $n$-graviton matrix element with a scalar loop. Representations for other spins in the loop
(Weyl fermion, vector, gravitino, graviton) can again be obtained from this by certain loop replacement rules
that are essentially independent applications of the above-mentioned QCD rules to the left- and right-mover parts, 
with an additional substitution rule $H_{ij} \longrightarrow 2/T$ for the cross terms.

\subsection{Symmetric partial integration}

Returning to the gluon case, the integrand resulting from the integration-by-parts procedure is unique for the two- and three-gluon cases, 
but starting from $n=4$ ambiguities appear \cite{Strassler:1992nc}.
Different algorithms lead to different equivalent integrands that are all free of $\ddot G_{ij}$s and suitable for an application of the pinching and loop
replacement rules.  Considering that the master formula \eqref{gluonmaster} possesses (apart from the colour ordering)
manifest permutation (or Bose) symmetry between
the $N$ gluons, in \cite{Schubert:1997ph} the following \emph{symmetric partial integration} algorithm was proposed that preserves this symmetry 
at each step. 

\begin{enumerate}

\item
In every step, partially integrate away {\sl all}
$\ddot G_{ij}$s appearing in the
term under inspection {\sl simultaneously}.
This is possible since different $\ddot G_{ij}$s never
share variables. 

\item
In the first step, for every $\ddot G_{ij}$ partially
integrate both over $\tau_i$ and $\tau_j$,
and take the mean of the results.

\item
At every following step, any $\ddot G_{ij}$
appearing must have been created in the 
previous step. Therefore either both $i$ and $j$
were used in the previous step, or just
one of them. If both, the rule is to again use both
variables in the actual step for partial integration,
and take the mean of the results. If only one of them
was used
in the previous step, 
then the other one should be used in the actual
step.

\end{enumerate}

This algorithm transforms the polynomial $P_n(\dot G_{ij},\ddot G_{ij})$ into a polynomial $Q_n(\dot G_{ij})$ that, unlike
$P_n$, is homogeneous not only in the polarisations, but also in the momenta. Together with the manifest permutation
invariance, this makes it possible to write $Q_n$ extremely compactly using a decomposition into bicycles and tails. 
A bicycle of length $k$ is defined by
\bear
\dot G(i_1,i_2,\cdots ,i_k) \equiv \dot G_{i_1i_2} 
\dot G_{i_2i_3} 
\cdots
\dot G_{i_ki_1}
Z_k(i_1,i_2,\dots ,i_k) 
\label{defbicycle}
\ear
where
\bear
Z_k(i_1,i_2,\dots ,i_k) \equiv \Bigl(\frac{1}{2}\Bigr)^{\delta_{k2}} {\rm tr}(f_{i_1}\cdots f_{i_k})
\ear
and $f_i^{\mu\nu} = k_i^\mu \varepsilon_i^{\nu} - \varepsilon_i^\mu k_i^\nu$ is the gluon field strength tensor. 
%These cycles are invariant under the operations of cyclic permutations and
%inversion. 
%For example, for $n=4$ there are only three really different cycles, which can be chosen as $\dot G(1234)$, $\dot G(1243)$, and $\dot G(1324)$. 
The tails are the left-overs after factorizing out all possible bicycles. The $k$-tail $T(i_1,i_2,\cdots ,i_k)$ involves $k$ polarisation vectors that
have not yet been absorbed into field strength tensors (it is possible to do so using further partial integrations with non-local coefficients \cite{Ahmadiniaz:2012ie},
but here we will not follow this route). 
%Although the tails are not manifestly transversal, they turn into total derivatives whenever any of the $\varepsilon_{i_m}$
%contained in them is replaced by $k_{i_m}$. 

For example, the cycle decompositions of $Q_3$ and $Q_4$ read
\begin{align}\label{Q3}
\begin{split}
Q_3&=Q_3^3+Q_3^2\\
Q_3^3 &=
\dot G(1,2,3)
\\
Q_3^2 &=
\dot G(1,2)T(3)+\dot G(2,3)T(1)+\dot G(3,1)T(2)
\end{split}
\end{align}
\begin{align}\label{Q4}
\begin{split}
Q_4 &= Q_4^4+Q_4^3+Q_4^2+Q_4^{22}\\
Q_4^4 &= \dot G(1,2,3,4)+\dot G(1,2,4,3) + \dot G(1,3,2,4)\\
Q_4^3 &= \dot G(1,2,3)T(4)+\dot G(2,3,4)T(1)+\dot G(3,4,1)T(2)+\dot G(4,1,2)T(3)\\
Q_4^2 &= \dot G(1,2)T(3,4)+\dot G(1,3)T(2,4)+\dot G(1,4)T(2,3)\\
 &\hspace{.3cm}+\dot G(2,3)T(1,4)+\dot G(2,4)T(1,3)+\dot G(3,4)T(1,2)\\
Q_4^{22} &=\dot G(1,2)\dot G(3,4)+\dot G(1,3)\dot G(2,4)+\dot G(1,4)\dot G(2,3)
\end{split}
\end{align}
the superscripts on the left-hand side indicating the cycle-content of a term.
Here the one- and two-tails appear,
\bear
T(a) &\equiv& \sum_r\dot G_{ar} \varepsilon_a\cdot k_r  
\label{defT1}\\
T(a,b) &\equiv&
\sum_{{r,s}\atop {(r,s)\ne (b,a)}}
\dot G_{ar}
\varepsilon_a\cdot k_r
\dot G_{s}
\varepsilon_b\cdot k_s
+\half
\dot G_{ab}
\varepsilon_a\cdot\varepsilon_b
\Bigl[
\sum_{r\ne b}
\dot G_{ar}k_a\cdot k_r 
- \sum_{s\ne a} \dot G_{s}k_b\cdot k_s
\Bigr]
\label{T2}
\ear
%In the two-tail, note the exclusion of terms from the sums that would correspond to the appearance of a new cycle $\dot G(a,b)$ in the tail, and thus to an overcounting. 
%Note that when advancing from the $n$-gluon amplitude to the $n+1$-gluon one the only new ingredient to be calculated is the $n-1$ tail, a relatively easy task. 
Note that the cycle decomposition of $Q_N$ involves the tails of length up to $N-2$. Up to length 4 the tails are given in
\cite{Schubert2001-73}; the five-tail was computed in \cite{Ahmadiniaz:2021fey}. 

For the graviton amplitudes, the above partial integration rules imply that the symmetric partial integration algorithm can be
applied separately in the variables $\tau_i$ to remove the $\ddot G_{ij}$ and in the $\bar\tau_i$ to remove the $\ddot {\bar G}_{ij}$,
with additional terms involving $H_{ij}$ generated by the first two rules in \eqref{crossterms}. The integrand can thus be ordered
according to the powers of $H_{ij}$, where the terms in the prefactor polynomial not containing any $H_{ij}$ can be factorised into
$Q_n(\dot {\bar G})Q_n(\dot G)$ and
terms with $m$ factors of $H$ containing $(n-m)$ factors of $\dot G$ and $\dot {\bar G}$ each. 

In this factorised term $Q_n(\dot {\bar G})Q_n(\dot G)$ we can apply all that we learned in the gluonic case in \cite{Ahmadiniaz:2021fey}. One of the lessons there was that we have to pinch only the tails in order to extract the multi-particle polarisations, that happen to obey colour-kinematics duality. Now we will have ``squared'' structures in the integrands like
\bear\label{eq:Dtail}
\bar{T}(1,2,\dots,n-2) T(1,2,\dots,n-2)
\ear
where we use the shorthand notation $\bar{T}(1,2,\dots,n-2)$ to imply that it only depends on $\dot {\bar G}$'s.

The idea now is to apply the procedure we implemented in \cite{Ahmadiniaz:2021fey} to calculate the Berends-Giele currents from the pinching of the squared tails to the gravitational case with this new structure in \eqref{eq:Dtail}. In the next section we will review this procedure for gluons before going back to the gravity version.

%For example, for $n=2$ the prefactor polynomial obtained by the expansion of the master exponent in \eqref{mastergrav}
%reads
%\bear
%&&\Bigl[k_1\c\pol_2 k_2\c\pol_1 \, \dot G_{12}^2 + \pol_1\c\pol_2\, \ddot G_{12} \Bigr ]
%  \Bigl[k_1\c\bar\pol_2 k_2\c\bar\pol_1 \, \dot {\bar G}^2_{12}
%         + \bar\pol_1\c\bar\pol_2\, \ddot {\bar G}_{12} \Bigr ]\nonumber\\
%&& +  \Bigl[ \pol_1\c\bar\pol_2 k_1\c\pol_2  k_2\c\bar\pol_1 +  \pol_2\c\bar\pol_1 k_2\c\pol_1  k_1\c\bar\pol_2 \Bigr] H_{12} \, \Gp{1}{2}\, \Gpb_{12}
%\nonumber\\
% && +  \pol_1\c\bar\pol_2  \pol_2\c\bar\pol_1 H_{12}^2  
% \label{Kred}
%\ear
%Following the above integration-by-parts rules this turns into
%
%\bear
%&& \Bigl[ k_1\c\pol_2 k_2\c\pol_1  - \pol_1\c\pol_2 k_1\c k_2\Bigr ]
%  \Bigl[ k_1\c\bar\pol_2 k_2\c\bar\pol_1  - \bar\pol_1\c\bar\pol_2 k_1\c k_2\Bigr ]
%  \, \dot G_{12}^2 \, \dot {\bar G}_{12}^2 
%         \nonumber\\
%&& 
% +  \Bigl[ \pol_1\c\bar\pol_2 k_1\c\pol_2  k_2\c\bar\pol_1 +  \pol_2\c\bar\pol_1 k_2\c\pol_1  k_1\c\bar\pol_2 \Bigr] H_{12} \, \Gp{1}{2}\, \Gpb_{12}
%\nonumber\\
% && + \Bigl[\pol_1\c\pol_2\bar\pol_1\c\bar\pol_2k_1\c k_2 -\pol_1\c\pol_2  k_1\c\bar\pol_2 k_2\c\bar\pol_1 -  \bar\pol_1\c\bar\pol_2  k_1\c\pol_2 k_2\c\pol_1 \Bigr] H_{12} \, \Gp{1}{2}\, \Gpb_{12}
% \nonumber\\
% && +  \pol_1\c\bar\pol_2  \pol_2\c\bar\pol_1 H_{12}^2  
% \label{KredIBP}
%\ear

%%%%%%%%%%%%%%%%%%%%%%%%%%%%%%%%%%%%%%%%%%%%%%%%%%%%%%%%%%%%%%%%%%%%%%%%%%%%%%%%%%%%

\section{Colour-kinematics duality}\label{sec:CK}
In the present section we will investigate how the colour-kinematics duality can be made manifest at the level of Yang-Mills Berends-Giele currents. We begin with a short summary of the results obtained in \cite{Ahmadiniaz:2021fey}, where a systematic procedure to construct such currents in the BCJ gauge from the Bern-Kosower formalism for one-loop gluon amplitudes is proposed. 

\subsection{Colour-stripped Berends-Giele currents from the Bern-Kosower formalism}
The main ingredients for the approach advocated in \cite{Ahmadiniaz:2021fey} are the symmetric partial integration algorithm and the Bern-Kosower rules reviewed in the preceding section. These two combine to produce the permutation invariant integrand $Q_n(\dot{G})$ and, for two adjacent legs $i$ and $j$ with $i < j$, the ``pinch operator'' acting on $Q_n(\dot{G})$ as
\begin{equation}\label{eq:pinchop}
\Dscr_{ij} Q_n(\dot{G}) = \frac{\partial}{\partial \dot{G}_{ij}} Q_{n}(\dot{G}) \bigg\vert_{\substack{\dot{G}_{ij}=0 \phantom{iiii} \\ \dot{G}_{jk} \rightarrow \dot{G}_{ik}} }. 
\end{equation}
The latter is what diagrammatically corresponds to pinching the two adjacent legs $i$ and $j$. Thus the complete effect of the pinching procedure in the Bern-Kosower formalism may be implemented by the iterated action of pinch operators. 

With the help of the foregoing we can build the colour-stripped Yang-Mills Berends-Giele currents directly in the BCJ gauge for both, the field strength and the polarisation. For the purposes of the present paper, however, 
we can restrict our attention to the latter. In doing so, we note --as pointed out in at the end of Section~IV of \cite{Ahmadiniaz:2021fey}-- that the part of the polynomial $Q_n(\dot{G})$ responsible for the extraction of the multi-particle fields associated with the colour-stripped Berends-Giele polarisation current, after applying the pinch operator consecutively $n-2$ times for each pair of labels following the cyclic order, is the $(n-2)$-tail $T(1,2,\dots,n-2)$. More explicitly, the exact meaning of this expression is
\begin{align}\label{eq:YMpinch-tail}
\begin{split}
\Dscr_{1(n-1)}\Dscr_{1(n-2)} \cdots \Dscr_{13} \Dscr_{12} T(1,2,\dots,n-2) = \varepsilon_{12 \cdots (n-2)} \cdot k_{n-1}. 
\end{split}
\end{align}
The multi-particle polarisation field $\varepsilon^{\mu}_{12 \cdots (n-2)}$ obtained this way satisfies the GJI of order $n-2$ in $1,2,\dots, n-2$. This property for these kind of polynomials has been verified up to degree $n=9$. For this reason, it is more accurate to write $\varepsilon^{\mu}_{\ell(12 \cdots (n-2))}$ instead of $\varepsilon^{\mu}_{12 \cdots (n-2)}$, or, more generally, $\varepsilon^{\mu}_{\ell(P)}$ instead of $\varepsilon^{\mu}_{P}$ for any word $P$.

Armed with these results, it is possible to find the explicit expression for the colour-stripped Berends-Giele polarisation current in terms of the multi-particle polarisation fields $\varepsilon^{\mu}_{P}$. To that end, it is very convenient to introduce a combinatorial artifact that helps us to keep track of the correspondence between nested Lie brackets and planar binary trees. This is termed the ``binary tree map'' in \cite{Bridges:2019siz}, but we will refer to it as the colour-stripped Berends-Giele map. It is defined as the map $b_{\cs}$ acting on all words and determined recursively by
\begin{align}\label{eq:csbm}
\begin{split}
b_{\cs}(i) &= i, \\
b_{\cs}(P) &= \frac{1}{s_P} \sum_{P =  QR} [b_{\cs}(Q),b_{\cs}(R)],
\end{split}
\end{align}
where $s_P$ is the Mandelstam invariant, and where $\sum_{P=QR}$ denotes the sum over all possible deconcatenations of the word $P$ into $Q$ and $R$. Also as a matter of notation, for an arbitrary labelled object $U_P$, such as the multi-particle polarisation fields $\varepsilon^{\mu}_P$, we bring the definition from \cite{Bridges:2019siz} for the replacement of words by such object as
\begin{equation}
\llbracket U \rrbracket \circ P = U_P. 
\end{equation}
With this background in mind, the colour-stripped Berends-Giele polarisation current associated with the multi-particle polarisation fields $\varepsilon^{\mu}_{P}$ is simply
\begin{equation}
\Acal^{\mu}_P = \llbracket \varepsilon^{\mu} \rrbracket \circ b_{\cs}(P). 
\end{equation}
We note, moreover, that the GJI satisfied by the multi-particle polarisation fields $\varepsilon^{\mu}_{P}$ translate directly into the shuffle symmetry $\Acal^{\mu}_{P \shuffle Q} = 0$. As an example, the colour-stripped Berends-Giele polarisation currents up to multiplicity four would read
\begin{align}
\begin{split}
\Acal^{\mu}_1 &= \varepsilon^{\mu}_1, \\
\Acal^{\mu}_{12} &= \frac{\varepsilon^{\mu}_{[1,2]}}{s_{12}}, \\
\Acal^{\mu}_{123} &= \frac{\varepsilon^{\mu}_{[[1,2],3]}}{s_{12}s_{123}} + \frac{\varepsilon^{\mu}_{[1,[2,3]]}}{s_{23}s_{123}}, \\
\Acal^{\mu}_{1234} &=  \frac{\varepsilon^{\mu}_{[[[1,2],3],4]}}{s_{12}s_{123}s_{1234}} + \frac{\varepsilon^{\mu}_{[[1,[2,3]],4]}}{s_{123}s_{1234}s_{23}} + \frac{\varepsilon^{\mu}_{[[1,2],[3,4]]}}{s_{12}s_{1234}s_{34}} +  \frac{\varepsilon^{\mu}_{[1,[[2,3],4]]}}{s_{1234}s_{23}s_{234}} +  \frac{\varepsilon^{\mu}_{[1,[2,[3,4]]]}}{s_{1234}s_{234}s_{34}}.
\end{split}
\end{align}
In these expressions, the multi-particle polarisation fields $\varepsilon^{\mu}_{[1,2]}$, $\varepsilon^{\mu}_{[[1,2],3]}$ and $\varepsilon^{\mu}_{[[[1,2],3],4]}$ can be written in a compact way as
\begin{align}\label{eq:Exmultiparticlepolarisations}
\begin{split}
\varepsilon^{\mu}_{[1,2]} &= \tfrac{1}{2} \big\{ (k_1 \cdot \varepsilon_2 )  \varepsilon_1^{\mu}  - (k_2 \cdot \varepsilon_1 ) \varepsilon_2^{\mu} + \varepsilon_{1 \nu} f_2^{\mu\nu} - \varepsilon_{2 \nu} f_1^{\mu\nu} \big\},\\
\varepsilon^{\mu}_{[[1,2],3]} &= \tfrac{1}{2} \big\{(k_3 \cdot \varepsilon_{[1,2]}) \varepsilon_3^{\mu} - (k_{12} \cdot \varepsilon_{3}) \varepsilon_{[1,2]}^{\mu} + \varepsilon_{3 \nu} f_{[1,2]}^{\mu\nu} - \varepsilon_{[1,2] \nu} f_{3}^{\mu\nu}  \big\} - k^{\mu}_{123} h_{123}, \\
\varepsilon^{\mu}_{[[[1,2],3],4]} &=  \tfrac{1}{2} \big\{(k_4 \cdot \varepsilon_{[[1,2],3]}) \varepsilon_4^{\mu} - (k_{123} \cdot \varepsilon_{4}) \varepsilon_{[[1,2],3]}^{\mu} + \varepsilon_{4 \nu} f_{[[1,2],3]}^{\mu\nu} - \varepsilon_{[[1,2],3] \nu} f_{4}^{\mu\nu}  \big\} + (k_{12} \cdot k_3) \varepsilon_3^{\mu} h_{124} \\
&\quad \quad\, + (k_1 \cdot k_2) (\varepsilon^{\mu}_2 h_{134} - \varepsilon^{\mu}_1 h_{234}) - k^{\mu}_{1234} h_{1234},
\end{split}
\end{align}
where we have set
\begin{align}\label{eq:Exmultiparticlestrength} 
\begin{split}
f^{\mu\nu}_i &= k^{\mu}_i \varepsilon^{\nu}_i - k^{\nu}_i \varepsilon^{\mu}_i, \\
f^{\mu\nu}_{[1,2]} &= k^{\mu}_{12} \varepsilon^{\nu}_{[1,2]} - k^{\nu}_{12} \varepsilon^{\mu}_{[1,2]} - (k_1\cdot k_2) (\varepsilon^{\mu}_1 \varepsilon^{\nu}_2 - \varepsilon^{\nu}_1 \varepsilon^{\mu}_2), \\
f^{\mu\nu}_{[[1,2],3]} &= k^{\mu}_{123} \varepsilon^{\nu}_{[[1,2],3]} - k^{\nu}_{123} \varepsilon^{\mu}_{[[1,2],3]} - (k_{12} \cdot k_3)(\varepsilon^{\mu}_{[1,2]} \varepsilon^{\nu}_3 - \varepsilon^{\nu}_{[1,2]} \varepsilon^{\mu}_3) - (k_1 \cdot k_2) (\varepsilon^{\mu}_{1} \varepsilon^{\nu}_{[2,3]} + \varepsilon^{\mu}_{[1,3]} \varepsilon^{\nu}_{2})  \\
&\quad\,   + (k_1 \cdot k_2) ( \varepsilon^{\nu}_{1} \varepsilon^{\mu}_{[2,3]} - \varepsilon^{\nu}_{[1,3]} \varepsilon^{\mu}_{2}) \\
h_{123} &= \tfrac{1}{4} (\varepsilon_1 \cdot \varepsilon_2) \varepsilon_3 \cdot (k_2 - k_1), \\
h_{1234} &= \tfrac{1}{4}\big[\varepsilon_1\cdot\varepsilon_2\,\varepsilon_3\cdot k_2 \varepsilon_4\cdot \,\left(k_1-k_{23}\right) + \tfrac{1}{2}\left(\varepsilon_1\cdot \varepsilon_2\,\varepsilon_3\cdot\varepsilon_4 \, k_2\cdot k_3   \right)-(123\rightarrow 312)\big] \,- (1\leftrightarrow 2). 
\end{split}
\end{align}

The bracketed notation in the words tell us about the GJI satisfied by the given object. Let us remind the GJI up to rank four, from \eqref{eq:GJI}, for the polarisations
\begin{align}\label{eq:GJI-pol} 
\begin{split}
&\varepsilon^{\mu}_{[1,2]} + \varepsilon^{\mu}_{[2,1]}=0,\\
&\varepsilon^{\mu}_{[[1,2],3]} + \varepsilon^{\mu}_{[[1,2],3]}=0,\quad\varepsilon^{\mu}_{[[1,2],3]}+\varepsilon^{\mu}_{[[3,1],2]}+\varepsilon^{\mu}_{[[2,3],1]}=0,\\
&\varepsilon^{\mu}_{[[1,2],3],4]} + \varepsilon^{\mu}_{[[1,2],3],4]}=0,\quad\varepsilon^{\mu}_{[[1,2],3],4]}+\varepsilon^{\mu}_{[[3,1],2],4]}+\varepsilon^{\mu}_{[[2,3],1],4]}=0,\\
&\varepsilon^{\mu}_{[[1,2],3],4]} - \varepsilon^{\mu}_{[[1,2],4],3]} + \varepsilon^{\mu}_{[[3,4],1],2]} - \varepsilon^{\mu}_{[[3,4],2],1]} = 0.
\end{split}
\end{align}

In addition, using the the symmetry properties of the bracket, $\varepsilon^{\mu}_{[[1,2],3]} = - \varepsilon^{\mu}_{[3,[1,2]]}$, $\varepsilon^{\mu}_{[[1,[2,3]],4]} =- \varepsilon^{\mu}_{[[[2,3],1],4]}$, $\varepsilon^{\mu}_{[1,[[2,3],4]]} = - \varepsilon^{\mu}_{[[[2,3],4],1]}$, $\varepsilon^{\mu}_{[1,[2,[3,4]]]} = \varepsilon^{\mu}_{[[[3,4],2],1]}$ and $\varepsilon^{\mu}_{[[1,2],[3,4]]} = \varepsilon^{\mu}_{[[[1,2],3],4]} -  \varepsilon^{\mu}_{[[[1,2],4],3]}$, so that these multi-particle polarisation fields are obtained from the formulas in \eqref{eq:Exmultiparticlepolarisations} by a simple relabelling. It is also worth pointing out that $f^{\mu\nu}_i$, $f^{\mu\nu}_{[1,2]}$ and $f^{\mu\nu}_{[[1,2],3]}$ in \eqref{eq:Exmultiparticlestrength} respectively correspond to the single-, two- and three-particle field strength produced following the procedure of \cite{Ahmadiniaz:2021fey}. 

Now that we have an explicit form of the the colour-stripped Berends-Giele polarisation currents, the next task is to write down the colour-stripped perturbiner expansion. This is a simple matter:~we just set it to be generating series 
\begin{equation}\label{eq:CS-pert}
A^{\mu}(x) = \sum_{n \geq 1} \sum_{P \in \Wscr_n} \Acal^{\mu}_P \ue^{\ui k_P \cdot x} T^{a_P},
\end{equation} 
where $\Wscr_n$ denotes the set of words of length $n$. It is important to note that the shuffle symmetry satisfied by the constituent currents $\Acal^{\mu}_P$ guarantees that the generating series \eqref{eq:CS-pert} is a Lie algebra-valued field. This expansion does not come directly from the Yang-Mills action, since in our case we have only trivalent vertices with no use of auxiliary fields. 

To complete our discussion we must also mention how the colour-stripped Berends-Giele polarisation currents $\Acal^{\mu}_P$ are related to the scattering amplitudes in Yang-Mills theory. At tree level, the colour-ordered partial amplitude of $n$ gluons is determined through the Berends-Giele formula
\begin{equation}\label{eq:cspartialamplitude}
\Ascr^{\mathrm{tree}}(1,2,\dots,n) = s_{12\cdots (n-1)} \Acal^{\mu}_{12\cdots (n-1)} \Acal_{n \mu}. 
\end{equation} 
The factor $s_{12\cdots (n-1)}$ is inserted to cancel the off-shell propagator inside $\Acal_{12\cdots (n-1)}$. Now that we are assuming momentum conservation and have on-shell external legs. There are other off-shell terms that cancel out, the ones of the form $k^{\mu}_Ph_P$ at the end of each polarisation. Finally it may be remarked that, by virtue of the shuffle symmetry, the partial amplitudes in the form of \eqref{eq:cspartialamplitude} satisfy the Kleiss-Kuijf relations \cite{Kleiss:1988ne}. 

\subsection{Colour-dressed Berends-Giele currents}
Now we turn our attention to obtaining the colour-dressed Berends-Giele polarisation currents from the multi-particle polarisation fields $\varepsilon_{P}^{\mu}$. These type of currents were obtained first for Yang-Mills in the Lorenz gauge in \cite{Mizera:2018jbh} using perturbiner methods. Back to our case, we need to make some small, but important, changes in the notation introduced thus far. In the first place, we need to modify the colour-stripped Berends-Giele map \eqref{eq:csbm} by a colour-dressed version of it, which we write as $b_{\cd}$. Here we borrow the prescription already encountered in~\cite{LAGQV21}. Namely, we define $b_{\cd}$ as the map acting on all ordered words and determined recursively by
\begin{align}\label{eq:cdBGmap}
\begin{split}
b_{\cd}(i) &= i, \\
b_{\cd}(P) &= \frac{1}{2 s_P} \sum_{P =  Q \cup R} [b_{\cd}(Q),b_{\cd}(R)],
\end{split}
\end{align}
where $\sum_{P =  Q \cup R}$ denotes the sum over all possible ways of distributing the letters of the ordered word $P$ into non-empty ordered words $Q$ and $R$. We  remark that 
the factor of $2$ in the denominator can be dropped if we impose the condition that $\lvert Q \rvert \geq \lvert R \rvert$. In the second place, for each ordered word $P = i_1 i_2 \cdots i_n$ of length $n$, we employ the notation $c_P^{a}$ to indicate the product of colour factors determined by
\begin{equation}\label{eq:cP}
c_P^{a} = \tilde{f}_{a_{i_1} a_{i_2}}^{\phantom{a_{i_1} a_{i_2}}b}  \tilde{f}_{b a_{i_3}}^{\phantom{b a_{i_3}}c}  \cdots  \tilde{f}_{d a_{i_{n-1}}}^{\phantom{d a_{i_{n-1}}}e} \tilde{f}_{e a_{i_n}}^{\phantom{e a_{i_n}}a},
\end{equation}
with the understanding that $c_i^{a} = \delta^{a}_{\phantom{a}a_i}$. We further put 
\begin{equation}\label{eq:c[P,Q]}
c_{[P,Q]}^{a} = \tilde{f}_{bc}^{\phantom{bc}a} c_P^{b} c_Q^{c}
\end{equation}
for any pair of ordered words $P$ and $Q$. In the third place, given two arbitrary labelled objects $U_{P}$ and $V_{P}$, we define the replacement of ordered words by the product of such objects as
\begin{equation}
\llbracket U \otimes V \rrbracket \circ P = U_P V_P. 
\end{equation}
By making use of the foregoing, one can show that we can write the colour-dressed Berends-Giele polarisation currents in the form
\begin{equation}\label{eq:cdBGcurrents}
\Acal_P^{a \mu} = \llbracket c^{a} \otimes \varepsilon^{\mu} \rrbracket \circ b_{\cd}(P). 
\end{equation}
At this point, however, we should perhaps emphasise that this way of representing the colour-dressed Berends-Giele polarisation currents is always possible regardless of whether or not the multi-particle polarisation fields $\varepsilon^{\mu}_P$ satisfy the GJI. When they do, as it is the case in the present discussion, we see that such identities mirror the GJI satisfied by the colour factor $c_P^{a}$. Hence, we are led to the conclusion that the ``factorisation'' of the colour-dressed Berends-Giele polarisation currents given in \eqref{eq:cdBGcurrents} is a realisation of the colour-kinematics duality. This will be pointed out, in a somewhat simplified context, and from a more algebraic perspective, in~\cite{LAGQV21}. In the next section, we will see that in terms of this factorisation, the double-copy prescription is straightforward to phrase.

We shall now proceed to write down explicitly the colour-dressed Berends-Giele polarisation currents up to multiplicity four, in order to familiarise ourselves with formula \eqref{eq:cdBGcurrents}. We first consider the single-particle case in which $P=1$. Then we at once obtain
\begin{equation}\label{eq:single-particle}
\Acal_1^{a\mu} = \delta^{a}_{\phantom{a} a_1} \varepsilon^{\mu}_{1}. 
\end{equation}
Next we consider the two-particle case in which $P =12$. In this case, the only possible way of distributing the letters is $(Q,R) = (1,2)$, and thus we find that colour-dressed Berends-Giele polarisation current $\Acal_{12}^{a\mu}$ acquires the form
\begin{equation}\label{eq:two-particle}
\Acal_{12}^{a\mu} = \frac{c^{a}_{[1,2]} \varepsilon^{\mu}_{[1,2]}}{s_{12}},
\end{equation}
with colour factor $c^{a}_{[1,2]} = \tilde{f}_{a_1 a_2}^{\phantom{a_1 a_2} a}$ and two-particle polarisation field $\varepsilon^{\mu}_{[1,2]}$ given by \eqref{eq:Exmultiparticlepolarisations}. Let us next take up the three-particle case in which $P = 123$. In this case, the possible ways of distributing the letters that contribute to the sum are $(Q,R) = (12,3), (13,2), (23,1)$. Therefore, after a straightforward calculation making use of the recursion \eqref{eq:cdBGmap} we obtain for the colour-dressed Berends-Giele polarisation current $\Acal_{123}^{a\mu}$ the formula
\begin{equation}\label{eq:three-particle}
\Acal_{123}^{a\mu} = \frac{c^{a}_{[[1,2],3]} \varepsilon^{\mu}_{[[1,2],3]}}{s_{12} s_{123}} +  \frac{c^{a}_{[[1,3],2]} \varepsilon^{\mu}_{[[1,3],2]}}{s_{13} s_{123}} +  \frac{c^{a}_{[[2,3],1]} \varepsilon^{\mu}_{[[2,3],1]}}{s_{23} s_{123}},
\end{equation}
with colour factors $c^{a}_{[[1,2],3]} = \tilde{f}_{a_1 a_2}^{\phantom{a_1 a_2} b}\tilde{f}_{b a_3}^{\phantom{b a_3} a}$, $c^{a}_{[[1,3],2]} = \tilde{f}_{a_1 a_3}^{\phantom{a_1 a_3} b}\tilde{f}_{b a_2}^{\phantom{b a_2} a}$, $c^{a}_{[[2,3],1]} = \tilde{f}_{a_2 a_3}^{\phantom{a_2 a_3} b}\tilde{f}_{b a_1}^{\phantom{b a_1} a}$ and three-particle polarisation fields $\varepsilon^{\mu}_{[[1,2],3]}$, $\varepsilon^{\mu}_{[[1,3],2]}$, $\varepsilon^{\mu}_{[[2,3],1]}$ given by \eqref{eq:Exmultiparticlepolarisations} after relabelling. Finally, we consider the four-particle case in which $P = 1234$. In this case, the possible ways of distributing the letters that contribute to the sum are $(Q,R)=(123,4),(124,3),(134,2),(234,1),(12,34),(13,24),(23,14)$. By analogy with the calculation leading to \eqref{eq:three-particle}, we find that the colour-dressed Berends-Giele polarisation current $\Acal_{1234}^{a\mu}$ may be represented in the form
\begin{align}\label{eq:four-particle}
\begin{split}
\Acal_{1234}^{a\mu} &= \frac{c^{a}_{[[[1,2],3],4]} \varepsilon^{\mu}_{[[[1,2],3],4]}}{s_{12} s_{123}s_{1234}}  + \frac{c^{a}_{[[[1,2],4],3]} \varepsilon^{\mu}_{[[[1,2],4],3]}}{s_{12} s_{124}s_{1234}} + \frac{c^{a}_{[[[1,3],4],2]} \varepsilon^{\mu}_{[[[1,3],4],2]}}{s_{13} s_{134}s_{1234}} + \frac{c^{a}_{[[[2,3],4],1]} \varepsilon^{\mu}_{[[[2,3],4],1]}}{s_{23} s_{234}s_{1234}}\\
&\quad\,    +  \frac{c^{a}_{[[[1,3],2],4]} \varepsilon^{\mu}_{[[[1,3],2],4]}}{s_{13} s_{123}s_{1234}} + \frac{c^{a}_{[[[1,4],2],3]} \varepsilon^{\mu}_{[[[1,4],2],3]}}{s_{14} s_{124}s_{1234}} + \frac{c^{a}_{[[[1,4],3],2]} \varepsilon^{\mu}_{[[[1,4],3],2]}}{s_{14} s_{134}s_{1234}} +  \frac{c^{a}_{[[[2,3],1],4]} \varepsilon^{\mu}_{[[[2,3],1],4]}}{s_{23} s_{123}s_{1234}}  \\
&\quad\,    + \frac{c^{a}_{[[[2,4],1],3]} \varepsilon^{\mu}_{[[[2,4],1],3]}}{s_{24} s_{124}s_{1234}}  + \frac{c^{a}_{[[[2,4],3],1]} \varepsilon^{\mu}_{[[[2,4],3],1]}}{s_{24} s_{234}s_{1234}} + \frac{c^{a}_{[[[3,4],1],2]} \varepsilon^{\mu}_{[[[3,4],1],2]}}{s_{34} s_{134}s_{1234}} +  \frac{c^{a}_{[[[3,4],2],1]} \varepsilon^{\mu}_{[[[3,4],2],1]}}{s_{34} s_{234}s_{1234}} \\
&\quad\,    + \frac{c^{a}_{[[1,2],[3,4]]} \varepsilon^{\mu}_{[[1,2],[3,4]]}}{s_{12} s_{34}s_{1234}} +  \frac{c^{a}_{[[1,3],[2,4]]} \varepsilon^{\mu}_{[[1,3],[2,4]]}}{s_{13} s_{24}s_{1234}} +  \frac{c^{a}_{[[1,4],[2,3]]} \varepsilon^{\mu}_{[[1,4],[2,3]]}}{s_{14} s_{23}s_{1234}}.
\end{split}
\end{align}
Here the colour factors are easily determined from \eqref{eq:cP} and \eqref{eq:c[P,Q]} as
\begin{alignat}{3}
c^{a}_{[[[1,2],3],4]} & = \tilde{f}^{\phantom{a_{1}a_{2}}b}_{a_{1}a_{2}}\tilde{f}^{\phantom{ba_{3}}c}_{ba_{3}}\tilde{f}^{\phantom{ca_{4}}a}_{ca_{4}},  &\qquad  c^{a}_{[[[1,2],4],3]} & = \tilde{f}^{\phantom{a_{1}a_{2}}b}_{a_{1}a_{2}}\tilde{f}^{\phantom{ba_{4}}c}_{ba_{4}}\tilde{f}^{\phantom{ca_{3}}a}_{ca_{3}}, &\qquad  c^{a}_{[[[1,3],4],2]} & = \tilde{f}^{\phantom{a_{1}a_{3}}b}_{a_{1}a_{3}}\tilde{f}^{\phantom{ba_{4}}c}_{ba_{4}}\tilde{f}^{\phantom{ca_{2}}a}_{ca_{2}},\nonumber\\
c^{a}_{[[[2,3],4],1]} & = \tilde{f}^{\phantom{a_{2}a_{3}}b}_{a_{2}a_{3}}\tilde{f}^{\phantom{ba_{4}}c}_{ba_{4}}\tilde{f}^{\phantom{ca_{1}}a}_{ca_{1}}, & c^{a}_{[[[1,3],2],4]} & = \tilde{f}^{\phantom{a_{1}a_{3}}b}_{a_{1}a_{3}}\tilde{f}^{\phantom{ba_{2}}c}_{ba_{2}}\tilde{f}^{\phantom{ca_{4}}a}_{ca_{4}} , &  c^{a}_{[[[1,4],2],3]} & = \tilde{f}^{\phantom{a_{1}a_{4}}b}_{a_{1}a_{4}}\tilde{f}^{\phantom{ba_{2}}c}_{ba_{2}}\tilde{f}^{\phantom{ca_{3}}a}_{ca_{3}},\nonumber\\
c^{a}_{[[[1,4],3],2]} & = \tilde{f}^{\phantom{a_{1}a_{4}}b}_{a_{1}a_{4}}\tilde{f}^{\phantom{ba_{3}}c}_{ba_{3}}\tilde{f}^{\phantom{ca_{2}}a}_{ca_{2}}, & c^{a}_{[[[2,3],1],4]} & = \tilde{f}^{\phantom{a_{2}a_{3}}b}_{a_{2}a_{3}}\tilde{f}^{\phantom{ba_{1}}c}_{ba_{1}}\tilde{f}^{\phantom{ca_{4}}a}_{ca_{4}} , & c^{a}_{[[[2,4],1],3]} & = \tilde{f}^{\phantom{a_{2}a_{4}}b}_{a_{2}a_{4}}\tilde{f}^{\phantom{ba_{1}}c}_{ba_{1}}\tilde{f}^{\phantom{ca_{3}}a}_{ca_{3}} ,\nonumber\\
c^{a}_{[[[2,4],3],1]} & = \tilde{f}^{\phantom{a_{2}a_{4}}b}_{a_{2}a_{4}}\tilde{f}^{\phantom{ba_{3}}c}_{ba_{3}}\tilde{f}^{\phantom{ca_{1}}a}_{ca_{1}}, &  c^{a}_{[[[3,4],1],2]} & = \tilde{f}^{\phantom{a_{3}a_{4}}b}_{a_{3}a_{4}}\tilde{f}^{\phantom{ba_{1}}c}_{ba_{1}}\tilde{f}^{\phantom{ca_{2}}a}_{ca_{2}}, &  c^{a}_{[[[3,4],2],1]} & = \tilde{f}^{\phantom{a_{3}a_{4}}b}_{a_{3}a_{4}}\tilde{f}^{\phantom{ba_{2}}c}_{ba_{2}}\tilde{f}^{\phantom{ca_{1}}a}_{ca_{1}},\nonumber\\
c^{a}_{[[1,2],[3,4]]} & = \tilde{f}^{\phantom{a_{1}a_{2}}b}_{a_{1}a_{2}}\tilde{f}^{\phantom{a_{3}a_{4}}c}_{a_{3}a_{4}}\tilde{f}^{\phantom{bc}a}_{bc}, &  c^{a}_{[[1,3],[2,4]]} & = \tilde{f}^{\phantom{a_{1}a_{3}}b}_{a_{1}a_{3}}\tilde{f}^{\phantom{a_{2}a_{4}}c}_{a_{2}a_{4}}\tilde{f}^{\phantom{bc}a}_{bc}, &  c^{a}_{[[1,4],[2,3]]} & = \tilde{f}^{\phantom{a_{1}a_{4}}b}_{a_{1}a_{4}}\tilde{f}^{\phantom{a_{2}a_{3}}c}_{a_{2}a_{3}}\tilde{f}^{\phantom{bc}a}_{bc}.\nonumber\\
\end{alignat}
As for the four-particle polarisation fields, keeping in mind the identities $\varepsilon^{\mu}_{[[1,2],[3,4]]} = \varepsilon^{\mu}_{[[[1,2],3],4]} -  \varepsilon^{\mu}_{[[[1,2],4],3]}$, $\varepsilon^{\mu}_{[[1,3],[2,4]]} = \varepsilon^{\mu}_{[[[1,3],2],4]} -  \varepsilon^{\mu}_{[[[1,3],4],2]}$ and $\varepsilon^{\mu}_{[[1,4],[2,3]]} = \varepsilon^{\mu}_{[[[1,4],2],3]} -  \varepsilon^{\mu}_{[[[1,4],3],2]}$, they are all determined by \eqref{eq:Exmultiparticlepolarisations} with the necessary relabellings.

Having obtained the expression \eqref{eq:cdBGcurrents} for the colour-dressed Berends-Giele polarisation currents, we can of course then obtain the colour-dressed perturbiner expansion. This is simply given as the generating series
\begin{equation}\label{eq:cdperturbiner}
A^{a\mu}(x) = \sum_{n \geq 1} \sum_{P \in \OWscr_n} \Acal^{a \mu}_P \ue^{\ui k_P \cdot x},
\end{equation}
where $\OWscr_n$ denotes the set of ordered words of length $n$. It should also be noted that the link between the colour-stripped and colour-dressed perturbiner expansions \eqref{eq:CS-pert} and \eqref{eq:cdperturbiner} is provided by $A^{\mu}(x) = A^{\mu}_{a}(x) T^{a}$. Finally we remark that a colour-dressed perturbiner expansion analogous to \eqref{eq:cdperturbiner} for the field strength can be obtained if we instead take the colour-dressed Berends-Giele field strength current associated with the multi-particle field strength. 

Before leaving this section, let us comment on the role the colour-dressed Berends-Giele polarisation currents $\Acal^{a\mu}_P$ play in the determination of the scattering amplitudes for Yang-Mills theory. Employing again the Berends-Giele formula we obtain the colour-dressed $n$-point amplitude
\begin{equation}\label{eq:cdpartialamplitude}
\Ascr^{\mathrm{tree}}_{n} = s_{12\cdots (n-1)} \Acal^{a\mu}_{12 \cdots (n-1)}  \Acal_{n a \mu} 
\end{equation}
where again we assume momentum conservation. It is also interesting to note that we may rewrite the amplitude \eqref{eq:cdpartialamplitude} as
\begin{equation}
\Ascr^{\mathrm{tree}}_{n} =  \sum_{\Gamma} \frac{c_{\Gamma} n_{\Gamma}}{\prod_{e \in \Gamma} s_e},
\end{equation}
where the sum goes over all $(2n-5)!!$ trivalent trees $\Gamma$ with propagators $s_e$ associated to each internal edge $e$ of $\Gamma$. Here $c_{\Gamma}$ denotes the colour structure attached to each diagram, while $n_{\Gamma}$ is the remaining part of the numerator involving kinematic information such as contractions of momenta and polarisation vectors. 

%%%%%%%%%%%%%%%%%%%%%%%%%%%%%%%%%%%%%%%%%%%%%%%%%%%%%%%%%%%%%%%%%%%%%%%%%%%%%%%%%%%%

\section{Double-copy relations for perturbiner expansions}\label{sec:DC}
In this section we will show that the double-copy prescription to construct gravity theories as the ``square'' of Yang-Mills theory finds a natural interpretation in terms of perturbiner expansions. To accomplish this, we first briefly discuss a systematic procedure, exactly analogous to the one found in \cite{Ahmadiniaz:2021fey}, to obtain the multi-particle polarisation fields on the gravity side from the Bern-Dunbar-Shimada formalism for one-loop graviton amplitudes. 

\subsection{Multi-particle polarisation tensors from the Bern-Dunbar-Shimada formalism}
We take as point of departure the symmetric partial integration algorithm and the Bern-Dunbar-Shimada rules explained in Section~\ref{sec:BK-BDS}. From these we identify the permutation invariant integrand $\bar{Q}_n(\dot{\bar{G}})Q_n(\dot{G})$. In addition, just as in Section~\ref{sec:CK} we have associated for two adjacent legs $i$ and $j$ with $i < j$ a pinch operator, we may likewise define a ``double pinch operator'' acting on $\bar{Q}_n(\dot{\bar{G}})Q_n(\dot{G})$ as
\begin{equation}
\bar{\Dscr}_{ij}\Dscr_{ij} \bar{Q}_n(\dot{\bar{G}})Q_n(\dot{G}) = \Bigg(\frac{\partial}{\partial \dot{\bar{G}}_{ij}} \bar{Q}_{n}(\dot{\bar{G}}) \bigg\vert_{\substack{\dot{\bar{G}}_{ij}=0 \phantom{iiii} \\ \dot{\bar{G}}_{jk} \rightarrow \dot{\bar{G}}_{ik}} }\Bigg)\Bigg(\frac{\partial}{\partial \dot{G}_{ij}} Q_{n}(\dot{G}) \bigg\vert_{\substack{\dot{G}_{ij}=0 \phantom{iiii} \\ \dot{G}_{jk} \rightarrow \dot{G}_{ik}} }\Bigg).
\end{equation}
This double pinch operator is thus identical with the one for Yang-Mills applied independently to both the left- and right-mover parts of the integrand expression. 

Our object is to find the multi-particle polarisation tensors by iterated action of double pinch operators. Here we may borrow from the analysis carried out in the Yang-Mills case, where we learned that the part of the polynomial $Q_n(\dot{G})$ relevant to the multi-particle polarisations is the $(n-2)$-tail. This makes it feasible in the present situation to also consider the $(n-2)$-tail $\bar{T}(1,2,\dots,n-2) T(1,2,\dots,n-2)$. Applying the double pinch operator consecutively $n-2$  times for each pair labels to the latter, one finds 
\begin{align}\label{eq:pichBDS}
\begin{split}
 & \bar{\Dscr}_{1(n-1)}\Dscr_{1(n-1)} \bar{\Dscr}_{1(n-2)}\Dscr_{1(n-2)}  \cdots \bar{\Dscr}_{13}\Dscr_{13}  \bar{\Dscr}_{12}\Dscr_{12} \bar{T}(1,2,\dots,n-2) T(1,2,\dots,n-2) \\
 &\qquad\qquad\qquad\qquad\qquad\qquad\qquad\qquad\qquad\qquad\qquad\qquad = \bar{\varepsilon}^{\mu}_{12 \cdots (n-2)} \varepsilon^{\nu}_{12 \cdots (n-2)} k_{(n-1)\mu}   k_{(n-1)\nu}.
\end{split}
\end{align}
This relation ensures that the multi-particle polarisation tensor is given by $\bar{\varepsilon}^{\mu}_{12 \cdots (n-2)} \varepsilon^{\nu}_{12 \cdots (n-2)}$. We may also remark that, by construction, each of the individual factors $\bar{\varepsilon}^{\mu}_{12 \cdots (n-2)}$ and $\varepsilon^{\nu}_{12 \cdots (n-2)}$ satisfies the generalised Jacobi identity of order $n-2$ in $1,2,\dots, n-2$. Thus this is precisely the ``square''  of the Yang-Mills multi-particle polarisation fields derived upon using  \eqref{eq:YMpinch-tail}. 

One further thing to be noted is this. In our preliminary discussion of the Bern-Dunbar-Shimada formalism, we indicated that when bringing into play the pinching rules we no longer have an ordering of the tree legs. This means that the tree attached to the loop is obtained by taking all possible pinches, which is an exceedingly tedious and onerous task. The main point to be stressed in connection with \eqref{eq:pichBDS} is that we may infer directly the existence of a ``double-copy'' version of the Berends-Giele polarisation currents, circumventing the need to determine them indirectly using the pinching procedure.

\subsection{Double-copy perturbiner expansion}\label{ssec:DCP}
The foregoing discussion contains all the underlying principles that are necessary for treating the double-copy polarisation currents and the corresponding perturbiner expansion. Indeed, examining the expression for the colour-dressed Berends-Giele polarisation current \eqref{eq:cdBGcurrents} and taking note of \eqref{eq:pichBDS} it is readily verified that the double-copy polarisation currents may be obtained by replacing the colour factor $\tilde{f}^{a}$ with another copy of the multi-particle polarisation field $\bar{\varepsilon}^{\mu}$. To be more precise, the double-copy polarisation current, which we denote by $\Gcal^{\mu\nu}_P$, is expressible in the form
\begin{equation}
\Gcal^{\mu\nu}_P = \llbracket \bar{\varepsilon}^{\mu} \otimes \varepsilon^{\nu} \rrbracket \circ b_{\cd}(P). 
\end{equation}
This provides a realisation of the off-shell double-copy that arises naturally in the string-based formalism, as an alternative to previous approaches \cite{Mizera:2018jbh,Wu:2021exa} that mimic the KLT relations adapting them to Berends-Giele currents.

As some examples, bringing to mind \eqref{eq:single-particle}, \eqref{eq:two-particle}, \eqref{eq:three-particle} and \eqref{eq:four-particle}, the first instances of the double-copy polarisation current up to multiplicity four are given by
\begin{align}\label{eq:DC-BG234}
\begin{split}
\Gcal^{\mu\nu}_{1} &= \bar{\varepsilon}^{\mu}_1 \varepsilon^{\nu}_{1}, \\
\Gcal_{12}^{\mu\nu} &= \frac{\bar{\varepsilon}^{\mu}_{[1,2]} \varepsilon^{\nu}_{[1,2]}}{s_{12}}, \\
\Gcal_{123}^{\mu\nu} &= \frac{\bar{\varepsilon}^{\mu}_{[[1,2],3]} \varepsilon^{\nu}_{[[1,2],3]}}{s_{12} s_{123}} +  \frac{\bar{\varepsilon}^{\mu}_{[[1,3],2]} \varepsilon^{\nu}_{[[1,3],2]}}{s_{13} s_{123}} +  \frac{\bar{\varepsilon}^{\mu}_{[[2,3],1]} \varepsilon^{\nu}_{[[2,3],1]}}{s_{23} s_{123}}, \\
\Gcal_{1234}^{\mu\nu} &= \frac{\bar{\varepsilon}^{\mu}_{[[[1,2],3],4]} \varepsilon^{\nu}_{[[[1,2],3],4]}}{s_{12} s_{123}s_{1234}}  + \frac{\bar{\varepsilon}^{\mu}_{[[[1,2],4],3]} \varepsilon^{\nu}_{[[[1,2],4],3]}}{s_{12} s_{124}s_{1234}} + \frac{\bar{\varepsilon}^{\mu}_{[[[1,3],4],2]} \varepsilon^{\nu}_{[[[1,3],4],2]}}{s_{13} s_{134}s_{1234}} + \frac{\bar{\varepsilon}^{\mu}_{[[[2,3],4],1]} \varepsilon^{\nu}_{[[[2,3],4],1]}}{s_{23} s_{234}s_{1234}}\\
&\quad\,    +  \frac{\bar{\varepsilon}^{\mu}_{[[[1,3],2],4]} \varepsilon^{\nu}_{[[[1,3],2],4]}}{s_{13} s_{123}s_{1234}} + \frac{\bar{\varepsilon}^{\mu}_{[[[1,4],2],3]} \varepsilon^{\nu}_{[[[1,4],2],3]}}{s_{14} s_{124}s_{1234}} + \frac{\bar{\varepsilon}^{\mu}_{[[[1,4],3],2]} \varepsilon^{\nu}_{[[[1,4],3],2]}}{s_{14} s_{134}s_{1234}} +  \frac{\bar{\varepsilon}^{\mu}_{[[[2,3],1],4]} \varepsilon^{\nu}_{[[[2,3],1],4]}}{s_{23} s_{123}s_{1234}}  \\
&\quad\,    + \frac{\bar{\varepsilon}^{\mu}_{[[[2,4],1],3]} \varepsilon^{\nu}_{[[[2,4],1],3]}}{s_{24} s_{124}s_{1234}}  + \frac{\bar{\varepsilon}^{\mu}_{[[[2,4],3],1]} \varepsilon^{\nu}_{[[[2,4],3],1]}}{s_{24} s_{234}s_{1234}} + \frac{\bar{\varepsilon}^{\mu}_{[[[3,4],1],2]} \varepsilon^{\nu}_{[[[3,4],1],2]}}{s_{34} s_{134}s_{1234}} +  \frac{\bar{\varepsilon}^{\mu}_{[[[3,4],2],1]} \varepsilon^{\nu}_{[[[3,4],2],1]}}{s_{34} s_{234}s_{1234}} \\
&\quad\,    + \frac{\bar{\varepsilon}^{\mu}_{[[1,2],[3,4]]} \varepsilon^{\nu}_{[[1,2],[3,4]]}}{s_{12} s_{34}s_{1234}} +  \frac{\bar{\varepsilon}^{\mu}_{[[1,3],[2,4]]} \varepsilon^{\nu}_{[[1,3],[2,4]]}}{s_{13} s_{24}s_{1234}} +  \frac{\bar{\varepsilon}^{\mu}_{[[1,4],[2,3]]} \varepsilon^{\nu}_{[[1,4],[2,3]]}}{s_{14} s_{23}s_{1234}}.
\end{split}
\end{align}
We reiterate that the crucial step in the double-copy procedure we have just argued is the construction of the multi-particle polarisation fields $\varepsilon^{\mu}_P$ satisfying the GJI as suggested by colour-kinematics duality. 

Since we have  already obtained the double-copy polarisation currents we can now readily obtain the double-copy perturbiner expansion, which is nothing but the generating series
\begin{equation}\label{eq:DCperturbiner}
G^{\mu\nu}(x) = \sum_{n \geq 1} \sum_{P \in \OWscr_n} \Gcal^{\mu\nu}_P \ue^{\ui k_P \cdot x}. 
\end{equation} 
Like in the Yang-Mills case, \eqref{eq:DCperturbiner} is not a solution of the Einstein field equations, for it has been ``strictified'' to include exclusively cubic interactions. 

Going on-shell now, it remains to say a word about the scattering amplitudes in the double-copy theory. Recalling the colour-dressed amplitude \eqref{eq:cdpartialamplitude}, the Berends-Giele formula for the $n$-point gravity amplitude reads
\begin{equation}\label{eq:BG-Amp}
\Mscr_n^{\mathrm{tree}} = s_{12\cdots (n-1)} \Gcal^{\mu\nu}_{12\cdots (n-1)} \Gcal_{n \mu\nu}. 
\end{equation}
Not surprisingly, the previous expression takes the well-known form for gravity amplitudes in its double-copy version
\begin{equation}
\Mscr^{\mathrm{tree}}_{n} =  \sum_{\Gamma} \frac{\bar{n}_{\Gamma} n_{\Gamma}}{\prod_{e \in \Gamma} s_e},
\end{equation}
which is equivalent to the KLT formula as can be seen in \cite{Bern:2019prr}. We also checked our result up to degree $n = 5$ for particular polarisations. At any rate, the outcome of this approach is that we can calculate the amplitudes for the double-copy theory in a relatively straightforward manner, without the need for
separately finding local BCJ numerators. This attribute was not apparent in previous approaches using the perturbiner method, since the generating series of Berends-Giele currents is usually presented in its colour-stripped version for the BCJ gauge.

%%%%%%%%%%%%%%%%%%%%%%%%%%%%%%%%%%%%%%%%%%%%%%%%%%%%%%%%%%%%%%%%%%%%%%%%%%%%%%%%%%

\section{Some other examples}\label{sec:examples}
Now that we found a prescription for the double-copy perturbiners, let us apply it to other theories beyond Yang-Mills and gravity. In principle it can be applied to any theory as soon as we guarantee multi-particle fields in the BCJ gauge. One first example should be the case where the BCJ gauge originally appeared, ten-dimensional $\mathscr{N}=1$ super Yang-Mills in \cite{Mafra:2014oia} (more recently from a new approach in \cite{Ben-Shahar:2021doh}), but for now we will restrict our presentation only to cases without supersymmetry.

\subsection{$\alpha'$-deformations}

For the first example we will calculate the currents and amplitudes for the deformations of general relativity that come from the $\alpha'$ corrections of the closed bosonic string, also referred to as GR+$R^2$+$R^3$. The amplitudes for this theory were calculated using the KLT relations for string theory \cite{Kawai:1985xq,Bjerrum-Bohr:2003hzh}. The action was found in \cite{Metsaev:1986yb} and it reads 
\begin{align}\label{eq:S-R2R3}
\begin{split}
S_{\substack{\mathrm{closed} \\ \mathrm{bosonic}}} & \sim  \int {\mathrm d}^D x \, \sqrt{g}  \left\{
R - 2(\partial_{\mu}\varphi)^2 - \frac{1}{12} H^2  + \frac{\alpha'}{4} e^{-2\varphi} \big( R_{\mu\nu\lambda\rho} R^{\mu\nu\lambda\rho} - 4 R_{\mu\nu} R^{\mu\nu}+R^2 \big) \right. \\
&\qquad\qquad\qquad\: +\alpha'^2  e^{-4\varphi}  \bigg( \frac{1}{16} R^{\mu \nu}{}_{\alpha\beta} R^{\alpha\beta}{}_{\lambda\rho} R^{\lambda\rho}{}_{\mu \nu}
- \frac{1}{12}R^{\mu\nu}{}_{\alpha\beta} R^{\nu\lambda}{}_{\beta\rho} R^{\lambda\mu}{}_{\rho \alpha} \bigg)
+ {\cal O}(\alpha'^3)  \Big\},
\end{split}
\end{align}
where here $\varphi$ represents the  dilaton and $H=dB$ represents the field strength of the $B$-field. The gauge field theory for the double-copy is the deformed Yang-Mills theory that comes from the low energy limit of the open string. The action, compatible with colour-kinematics duality \cite{Broedel:2012rc}, is the following 
\bear\label{eq:S-F3F4}
S_{\mathrm{YM}+F^3+F^4} = \int {\mathrm d}^D x \ \mathrm{tr} \left\{\frac{1}{4}F_{\mu \nu}F^{\mu \nu} + \frac{ 2\alpha'}{3}F_\mu{}^\nu F_\nu{}^\lambda F_{\lambda}{}^\mu + \frac{ \alpha'^2}{4}[F_{\mu \nu}, F_{\lambda\rho}]  [F^{\mu \nu},F^{\lambda\rho}]\right\},
\ear
which has the following equations of motion in the Lorenz gauge, $\partial_{\mu}A^{\mu} = 0$,
\bear\label{eq:EL-F3F4}
\Box A^{\lambda} &=& [A^\mu , \partial_{\mu} A^{\lambda}] + [A_{\mu} ,F^{\mu\lambda}] + 2\alpha' \big\{[\nabla_\mu F^{\mu\nu},F_\nu{}^\lambda] + [F^{\mu\nu},\nabla_\mu F_\nu{}^\lambda]\big\}\notag\\
&&+ 2\alpha'^2\Big\{\big[[\nabla_\mu F^{\mu\lambda},F_{\rho\sigma}],F^{\rho\sigma}\big]  
+ \big[[F^{\mu\lambda},\nabla_\mu F_{\rho\sigma}],F^{\rho\sigma}\big] + \big[[F^{\mu\lambda},F_{\rho\sigma}],\nabla_\mu F^{\rho\sigma}\big]\Big\}.
\ear
In \cite{Garozzo:2018uzj}, the authors conducted a detailed analysis for the calculation of the currents in this gauge using the perturbiner approach \cite{Mafra:2016ltu,Mizera:2018jbh,Lopez-Arcos:2019hvg}. Then, they applied the non-linear gauge transformation studied in \cite{Lee:2015upy} in order to obtain currents in the BCJ gauge. In general the expressions for the $\alpha'$-deformed multi-particle polarisations have the following structure
\bear
a^{\mu}_P=\varepsilon^{\mu}_P + \alpha'\varepsilon^{(1)\mu}_P + \alpha'^2\varepsilon^{(2)\mu}_P.
\ear
We invite the reader to have a look at the explicit expressions in \cite{Garozzo:2018uzj}.

Our double-copy perturbiner for the $\alpha'$-deformation of general relativity comes out to be
\bear
G^{(\alpha')\mu\nu}(x)=\sum_{n \geq 1} \sum_{P \in \mathcal{OW}_n}\mathcal{G}^{(\alpha')\mu\nu}_P \ue^{\mathrm{i}k_P\cdot x},
\ear
where the Berends-Giele currents is given by
\bear
\mathcal{G}^{(\alpha')\mu\nu}_P&=&\llbracket a^{\mu}\otimes\bar{a}^{\nu}  \rrbracket \circ b_{cd}(P).
\ear
Naturally, in complete analogy with \eqref{eq:BG-Amp}, the corresponding amplitude reads
\bear
\mathscr{M}_n^{(\alpha')\mathrm{tree}} &=& s_{1\dots n-1}\mathcal{G}_{1\dots n-1}^{(\alpha')\mu\nu}\mathcal{G}^{(\alpha')}_{n\mu\nu}.
\ear
This we have also checked using the explicit expressions for $a^{\mu}_P$ from \cite{Garozzo:2018uzj}.

\subsection{Zeroth-copy}
Another example whose perturbiner can be obtained in a very straightforward manner with our approach is the one for the bi-adjoint scalar model. For this model, originally found in \cite{Cachazo:2013iaa}, we have a scalar field that takes values in the tensor product $\mathfrak{su}(N) \otimes \mathfrak{su}(N')$, and is expressible in terms of the generators as $\Phi=\Phi_{aa'}T^a \otimes T'^{a'}$. The corresponding action takes the form
\bear
S_{\text{bi-adjoint}}=\int {\mathrm d}^D x \left\{-\tfrac{1}{2}\Phi^{aa'}\Box\Phi_{aa'} + \frac{1}{3!}\tilde{f}^{abc}\tilde{f'}^{a'b'c'}\Phi_{aa'}\Phi_{bb'}\Phi_{cc'}\right\}.
\ear
Its Berends-Giele currents were found for the first time in \cite{Mafra:2016ltu} in the colour-stripped version and the colour-dressed version in \cite{Mizera:2018jbh}, both cases using the perturbiner approach. Here we can obtain it simply by applying the zeroth-copy \cite{White:2020sfn}, now in its analogue perturbiner version. Therefore, for the bi-adjoint perturbiner we have
\bear
\Phi^{aa'}(x)=\sum_{n \geq 1} \sum_{P \in \mathcal{OW}_n}\phi^{aa'}_P \ue^{\mathrm{i}k_P\cdot x},
\ear
where for the Berends-Giele currents read
\bear
\phi^{aa'}_P=\llbracket c^{a}\otimes c'^{a'}\rrbracket\circ b_{cd}(P)
\ear 
The expressions for the currents are exactly like the ones in \eqref{eq:DC-BG234} but replacing the polarisations by the colour factors presented in Section \ref{sec:CK}. The colour-dressed amplitudes can also be calculated directly using the Berends-Giele formula in \eqref{eq:BG-Amp}.

%%%%%%%%%%%%%%%%%%%%%%%%%%%%%%%%%%%%%%%%%%%%%%%%%%%%%%%%%%%%%%%%%%%%%%%%%%%%%%%%%%%%

\section{Conclusions}\label{sec:concl}

We have seen that colour-kinematics duality and double-copy arise quite naturally in the string-based formalism in the form of multi-particle fields. The combination of the string-based rules and the technology developed in the study of such fields allowed us to reduce the calculation of tree-level amplitudes in Yang-Mills and gravity to a single basic calculation, namely the full pinching of the tail that gives the multi-particle polarisation in the BCJ gauge. 
Both the colour-dressed Yang-Mills Berends-Giele currents in the BJC gauge and the gravitational Berends-Giele
currents are given explicitly up to multiplicity five. 
The most attractive feature of our formalism is that it never becomes necessary to determine gauge transformation terms. 
We presented a new prescription for the off-shell double-copy that has applications to theories beyond the ones that we can represent by the infinite string tension limit, as we have demonstrated with the examples of
$\alpha'$-deformed gravity and the bi-adjoint scalar model. 

In the Yang-Mills case, we have shown in \cite{Ahmadiniaz:2021fey} how to feed the obtained multi-particle tensors
back into the Bern-Kosower formalism so as to make the whole pinching procedure unnecessary. 
It is not obvious whether this aspect of our approach can be generalized to the gravity case, since here the
existence of the cross terms seems to start making a real difference. We leave this to further study.  

One application in progress is the calculation of Berends-Giele currents for gravity coupled to matter fields along the lines of \cite{Gomez:2020vat} and \cite{Johansson:2019dnu}, that could be compared with \cite{Gomez:2021shh}. Another application for the near future is to some cases of supergravity, where Berends-Giele currents have been found for $\mathscr{N}=1$ Super-Yang-Mills in the BCJ gauge in \cite{Mafra:2014oia,Bridges:2019siz}.

\section*{Acknowledgements}

We thank Lucia Garozzo for providing us with the file with their $\alpha'$-deformed polarisations. CLA thanks Oliver Schlotterer for the very fruitful discussions. CS thanks Lance Dixon and Piotr Tourkine for various informations and
discussions.
FMB is supported in part by the European Research Council under ERC-STG-804286 UNISCAMP.

\appendix
\section{Berends-Giele currents of multiplicity five}\label{sec:appA}

In this appendix, we exploit the procedure showed above for the computation of colour-dressed Berends-Giele polarisation currents in the BCJ gauge and we show the complete expressions for the current at multiplicity five. The computation follows from (\ref{eq:cdBGcurrents}), where in the five-particle case $P=12345$. In this case, the word decomposition reads as $(Q,R)=(1234)(5)$, $(1235)(4)$, $(1245)(3)$, $(1345)(2)$, $(2345)(1)$, $(123)(45)$, $(124)(35)$, $(125)(34)$, $(134)(25)$, $(145)(23)$, $(135)(24)$, $(234)(15)$, $(235)(14)$, $(245)(13)$, $(345)(12)$. Therefore, making also use of the recursion in (3.11), we obtain for the colour-dressed Berends-Giele polarisation current $\mathcal{A}_{12345}^{a\mu}$ the formula
\begin{align}\label{eq:CD-BG5}
\begin{split}
\mathcal{A}_{12345}^{a\mu} &= \frac{c^{a}_{[[[[1,2],3],4],5]}\varepsilon^{\mu}_{[[[[1,2],3],4],5]}}{s_{12}s_{123}s_{1234}s_{12345}} + \frac{c^{a}_{[[[[1,3],2],4],5]}\varepsilon^{\mu}_{[[[[1,3],2],4],5]}}{s_{13}s_{123}s_{1234}s_{12345}} + \frac{c^{a}_{[[[[2,3],1],4],5]}\varepsilon^{\mu}_{[[[[2,3],1],4],5]}}{s_{23}s_{123}s_{1234}s_{12345}}\\ 
&+ \frac{c^{a}_{[[[[1,2],4],3],5]}\varepsilon^{\mu}_{[[[[1,2],4],3],5]}}{s_{12}s_{124}s_{1234}s_{12345}}+ \frac{c^{a}_{[[[[1,4],2],3],5]}\varepsilon^{\mu}_{[[[[1,4],2],3],5]}}{s_{14}s_{124}s_{1234}s_{12345}} + \frac{c^{a}_{[[[[2,4],1],3],5]}\varepsilon^{\mu}_{[[[[2,4],1],3],5]}}{s_{24}s_{124}s_{1234}s_{12345}}\\
&+ \frac{c^{a}_{[[[[1,3],4],2],5]}\varepsilon^{\mu}_{[[[[1,3],4],2],5]}}{s_{13}s_{134}s_{1234}s_{12345}} + \frac{c^{a}_{[[[[1,4],3],2],5]}\varepsilon^{\mu}_{[[[[1,4],3],2],5]}}{s_{14}s_{134}s_{1234}s_{12345}}+ \frac{c^{a}_{[[[[3,4],1],2],5]}\varepsilon^{\mu}_{[[[[3,4],1],2],5]}}{s_{34}s_{134}s_{1234}s_{12345}}\\
&+ \frac{c^{a}_{[[[[2,3],4],1],5]}\varepsilon^{\mu}_{[[[[2,3],4],1],5]}}{s_{23}s_{234}s_{1234}s_{12345}} + \frac{c^{a}_{[[[[2,4],3],1],5]}\varepsilon^{\mu}_{[[[[2,4],3],1],5]}}{s_{24}s_{234}s_{1234}s_{12345}} + \frac{c^{a}_{[[[[3,4],2],1],5]}\varepsilon^{\mu}_{[[[[3,4],2],1],5]}}{s_{34}s_{234}s_{1234}s_{12345}}\\
&+ \frac{c^{a}_{[[[1,2],[3,4]],5]}\varepsilon^{\mu}_{[[[1,2],[3,4]],5]}}{s_{12}s_{34}s_{1234}s_{12345}} + \frac{c^{a}_{[[[1,3],[2,4]],5]}\varepsilon^{\mu}_{[[[1,3],[2,4]],5]}}{s_{13}s_{24}s_{1234}s_{12345}} + \frac{c^{a}_{[[[1,4],[2,3]],5]}\varepsilon^{\mu}_{[[[1,4],[2,3]],5]}}{s_{14}s_{23}s_{1234}s_{12345}}\\
&+\Big( (1234)(5)\leftrightarrow (1235)(4)\Big)+\Big( (1234)(5)\leftrightarrow (1245)(3)\Big)\\
&+\Big( (1234)(5)\leftrightarrow (1345)(2)\Big)+\Big( (1234)(5)\leftrightarrow (2345)(1)\Big)\\
&+\frac{c^{a}_{[[[1,2],3],[4,5]]}\varepsilon^{\mu}_{[[[1,2],3],[4,5]]}}{s_{12}s_{123}s_{45}s_{12345}} + \frac{c^{a}_{[[[2,3],1],[4,5]]}\varepsilon^{\mu}_{[[[2,3],1],[4,5]]}}{s_{23}s_{123}s_{45}s_{12345}} + \frac{c^{a}_{[[[1,3],2],[4,5]]}\varepsilon^{\mu}_{[[[1,3],2],[4,5]]}}{s_{13}s_{123}s_{45}s_{12345}}\\
&+\Big( (123)(45)\leftrightarrow (124)(35)\Big)+\Big( (123)(45)\leftrightarrow (125)(34)\Big)+\Big( (123)(45)\leftrightarrow (134)(25)\Big)\\
&+\Big( (123)(45)\leftrightarrow (145)(23)\Big)+\Big( (123)(45)\leftrightarrow (135)(24)\Big)+\Big( (123)(45)\leftrightarrow (234)(15)\Big)\\
&+\Big( (123)(45)\leftrightarrow (235)(14)\Big)+\Big( (123)(45)\leftrightarrow (245)(13)\Big)+\Big( (123)(45)\leftrightarrow (345)(12)\Big),
\end{split}
\end{align}
where the colour factors have structures of type
\begin{alignat}{2}
\begin{split}
c^{a}_{[[[[1,2],3],4],5]}  = \tilde{f}^{\phantom{a_{1}a_{2}}b}_{a_{1}a_{2}}&\tilde{f}^{\phantom{ba_{3}}c}_{ba_{3}}\tilde{f}^{\phantom{ca_{4}}d}_{ca_{4}}\tilde{f}^{\phantom{da_{5}}a}_{da_{5}},  \qquad c^{a}_{[[[1,2],[3,4]],5]}= \tilde{f}^{\phantom{a_{1}a_{2}}b}_{a_{1}a_{2}}\tilde{f}^{\phantom{a_{3}a_{4}}c}_{a_{3}a_{4}}\tilde{f}^{\phantom{bc}d}_{bc}\tilde{f}^{\phantom{da_{5}}a}_{da_{5}},\\
& c^{a}_{[[[1,2],3],[4,5]]}= \tilde{f}^{\phantom{a_{1}a_{2}}b}_{a_{1}a_{2}}\tilde{f}^{\phantom{ba_{3}}c}_{ba_{3}}\tilde{f}^{\phantom{a_{4}a_{5}}d}_{a_{4}a_{5}}\tilde{f}^{\phantom{cd}a}_{cd}.
\end{split}
\end{alignat}
Also, in (\ref{eq:CD-BG5}) we made use of the multi-particle polarisation field $\varepsilon_{[[[[1,2],3],4],5]}^\mu$ defined as
\begin{align}\label{eq:e12345}
\begin{split}
\varepsilon_{[[[[1,2],3],4],5]}^\mu &= \frac{1}{2}\big[\varepsilon_5^\mu\left(\varepsilon_{[[[1,2],3],4]}\cdot k_5\right)-\varepsilon_{[[[1,2],3],4]}^\mu\left(\varepsilon_{5}\cdot k_{1234}\right)+\varepsilon_{[[[1,2],3],4]\nu}f_5^{\nu\mu}-\varepsilon_{5\nu}f_{[[[1,2],3],4]}^{\nu\mu}\big]\\
&+(k_{123}\cdot k_4)\varepsilon_4^{\mu}h_{1235}+ (k_{12}\cdot k_3)\big(\varepsilon_3^{\mu}h_{1245}+\varepsilon_{[3,4]}^\mu h_{125}-\varepsilon_{[1,2]}^\mu h_{345}\big)\\
&+(k_{1}\cdot k_2)\big(\varepsilon_2^\mu h_{1345} + \varepsilon_{[2,3]}^\mu h_{145} + \varepsilon_{[2,4]}^\mu h_{135} - \varepsilon_1^\mu h_{2345} - \varepsilon_{[1,3]}^\mu h_{245} - \varepsilon_{[1,4]}^\mu h_{235}\big)-k_{12345}^\mu h_{12345}
\end{split}
\end{align}
where we have set
\begin{align}
\begin{split}
f^{\mu\nu}_{[[[1,2],3],4]} &= k^{\mu}_{1234} \varepsilon^{\nu}_{[[[1,2],3],4]} - (k_{123} \cdot k_4)\varepsilon^{\mu}_{[[1,2],3]} \varepsilon^{\nu}_4  - (k_{12} \cdot k_3) (\varepsilon^{\mu}_{[1,2]} \varepsilon^{\nu}_{[3,4]} + \varepsilon^{\mu}_{[[1,2],4]} \varepsilon^{\nu}_{3})  \\
&  - (k_1 \cdot k_2) ( \varepsilon^{\mu}_{[1,3]} \varepsilon^{\nu}_{[2,4]}+\varepsilon^{\mu}_{[1,4]} \varepsilon^{\nu}_{[2,3]}+\varepsilon^{\mu}_{[[1,3],4]}\varepsilon^{\nu}_{2}-\varepsilon^{\mu}_{[[2,3],4]}\varepsilon^{\nu}_{1})-(\mu\leftrightarrow\nu)
\end{split}
\end{align}
and
\begin{align}
\begin{split}
h_{12345}&=\frac{1}{4}\bigg[\Big\lbrace\epseps34 \epsk12\epsk23\epsk51 - \epseps23\epsk12\epsk43\epsk51 + \epseps12\epsk32\epsk43\epsk51\\
&+ \epseps34\epsk12\epsk23\epsk52- \epseps23\epsk12\epsk43\epsk52 - \epseps12\epsk32\epsk43\epsk52\\
&- \epseps15\epseps34\epsk23k_1\cdot k_2 + \frac{1}{2}\epseps15\epseps23\epsk43k_1\cdot k_2- \frac{1}{2}\epseps25\epseps34\epsk12k_2\cdot k_3\\
& - \frac{3}{4}\epseps12\epseps34\epsk51k_2\cdot k_3 + \frac{3}{4}\epseps12\epseps34\epsk52k_2\cdot k_3- (123\rightarrow 312)- (1234\rightarrow 4123)\Big\rbrace\\
& + \Big\lbrace\epseps23\epsk12\epsk42\epsk53-\epseps24\epsk12\epsk32\epsk51- \epseps23\epsk12\epsk42\epsk51\\
&- \epseps12\epsk32\epsk42\epsk52 - \epseps12\epsk32\epsk42\epsk54 + \frac{1}{2}\epseps24\epseps35\epsk12k_2\cdot k_3\\
& + \frac{1}{2}\epseps12\epseps35\epsk42k_2\cdot k_3+ \frac{1}{2}\epseps12\epseps45\epsk32k_2\cdot k_4 - (123\rightarrow 312)\Big\rbrace\\
&+ \epseps14\epsk21\epsk31\epsk51 + \epseps24\epsk12\epsk32\epsk54 \bigg]- (1\rightarrow 2).
\end{split}
\end{align} 
Note that, as for the lower-point polarisation fields, $\varepsilon_{[[[[1,2],3],4],5]}^\mu$ is the only 
five-particle polarisation field needed in (\ref{eq:CD-BG5}). Indeed, using the identities $\varepsilon^{\mu}_{[[[1,2],[3,4]],5]}=\varepsilon^{\mu}_{[[[[1,2],3],4],5]}-\varepsilon^{\mu}_{[[[[1,2],4],3],5]}$ and $\varepsilon^{\mu}_{[[[1,2],3],[4,5]]}=\varepsilon^{\mu}_{[[[[1,2],3],4],5]}-\varepsilon^{\mu}_{[[[[1,2],3],5],4]}$, all the five-particle polarisation fields appearing in (\ref{eq:CD-BG5}) are obtained from (\ref{eq:e12345}) by a simple relabelling.\\

Finally, we report here the formula for the double-copy polarisation current at five point $\mathcal{G}_{12345}^{\mu\nu}$. As pointed out in Section \ref{sec:DC}, this is simply obtained by substituting the colour factors $c^a$ in (\ref{eq:CD-BG5}) with another copy of the multi-particle polarisation field $\bar\varepsilon^\mu$. The resulting expression for $\mathcal{G}_{12345}^{\mu\nu}$ is thus given by
\begin{align}\label{eq:DC-BG12345}
\begin{split}
\mathcal{G}_{12345}^{\mu\nu} &= \frac{\varepsilon^{\mu}_{[[[[1,2],3],4],5]}\bar\varepsilon^{\nu}_{[[[[1,2],3],4],5]}}{s_{12}s_{123}s_{1234}s_{12345}} + \frac{\varepsilon^{\mu}_{[[[[1,3],2],4],5]}\bar\varepsilon^{\nu}_{[[[[1,3],2],4],5]}}{s_{13}s_{123}s_{1234}s_{12345}} + \frac{\varepsilon^{\mu}_{[[[[2,3],1],4],5]}\bar\varepsilon^{\nu}_{[[[[2,3],1],4],5]}}{s_{23}s_{123}s_{1234}s_{12345}}\\
&+ \frac{\varepsilon^{\mu}_{[[[[1,2],4],3],5]}\bar\varepsilon^{\nu}_{[[[[1,2],4],3],5]}}{s_{12}s_{124}s_{1234}s_{12345}}+ \frac{\varepsilon^{\mu}_{[[[[1,4],2],3],5]}\bar\varepsilon^{\nu}_{[[[[1,4],2],3],5]}}{s_{14}s_{124}s_{1234}s_{12345}} + \frac{\varepsilon^{\mu}_{[[[[2,4],1],3],5]}\bar\varepsilon^{\nu}_{[[[[2,4],1],3],5]}}{s_{24}s_{124}s_{1234}s_{12345}}\\
&+ \frac{\varepsilon^{\mu}_{[[[[1,3],4],2],5]}\bar\varepsilon^{\nu}_{[[[[1,3],4],2],5]}}{s_{13}s_{134}s_{1234}s_{12345}} + \frac{\varepsilon^{\mu}_{[[[[1,4],3],2],5]}\bar\varepsilon^{\nu}_{[[[[1,4],3],2],5]}}{s_{14}s_{134}s_{1234}s_{12345}}+ \frac{\varepsilon^{\mu}_{[[[[3,4],1],2],5]}\bar\varepsilon^{\nu}_{[[[[3,4],1],2],5]}}{s_{34}s_{134}s_{1234}s_{12345}}\\
&+ \frac{\varepsilon^{\mu}_{[[[[2,3],4],1],5]}\bar\varepsilon^{\nu}_{[[[[2,3],4],1],5]}}{s_{23}s_{234}s_{1234}s_{12345}} + \frac{\varepsilon^{\mu}_{[[[[2,4],3],1],5]}\bar\varepsilon^{\nu}_{[[[[2,4],3],1],5]}}{s_{24}s_{234}s_{1234}s_{12345}} + \frac{\varepsilon^{\mu}_{[[[[3,4],2],1],5]}\bar\varepsilon^{\nu}_{[[[[3,4],2],1],5]}}{s_{34}s_{234}s_{1234}s_{12345}}\\
&+ \frac{\varepsilon^{\mu}_{[[[1,2],[3,4]],5]}\bar\varepsilon^{\nu}_{[[[1,2],[3,4]],5]}}{s_{12}s_{34}s_{1234}s_{12345}} + \frac{\varepsilon^{\mu}_{[[[1,3],[2,4]],5]}\bar\varepsilon^{\nu}_{[[[1,3],[2,4]],5]}}{s_{13}s_{24}s_{1234}s_{12345}} + \frac{\varepsilon^{\mu}_{[[[1,4],[2,3]],5]}\bar\varepsilon^{\nu}_{[[[1,4],[2,3]],5]}}{s_{14}s_{23}s_{1234}s_{12345}}\\
&+\Big( (1234)(5)\leftrightarrow (1235)(4)\Big)+\Big( (1234)(5)\leftrightarrow (1245)(3)\Big)\\
&+\Big( (1234)(5)\leftrightarrow (1345)(2)\Big)+\Big( (1234)(5)\leftrightarrow (2345)(1)\Big)\\
&+\frac{\varepsilon^{\mu}_{[[[1,2],3],[4,5]]}\bar\varepsilon^{\nu}_{[[[1,2],3],[4,5]]}}{s_{12}s_{123}s_{45}s_{12345}} + \frac{\varepsilon^{\mu}_{[[[2,3],1],[4,5]]}\bar\varepsilon^{\nu}_{[[[2,3],1],[4,5]]}}{s_{23}s_{123}s_{45}s_{12345}} + \frac{\varepsilon^{\mu}_{[[[1,3],2],[4,5]]}\bar\varepsilon^{\nu}_{[[[1,3],2],[4,5]]}}{s_{13}s_{123}s_{45}s_{12345}}\\
&+\Big( (123)(45)\leftrightarrow (124)(35)\Big)+\Big( (123)(45)\leftrightarrow (125)(34)\Big)+\Big( (123)(45)\leftrightarrow (134)(25)\Big)\\
&+\Big( (123)(45)\leftrightarrow (145)(23)\Big)+\Big( (123)(45)\leftrightarrow (135)(24)\Big)+\Big( (123)(45)\leftrightarrow (234)(15)\Big)\\
&+\Big( (123)(45)\leftrightarrow (235)(14)\Big)+\Big( (123)(45)\leftrightarrow (245)(13)\Big)+\Big( (123)(45)\leftrightarrow (345)(12)\Big).
\end{split}
\end{align}


\section*{References}

\begin{thebibliography}{10}
\expandafter\ifx\csname url\endcsname\relax
  \def\url#1{\texttt{#1}}\fi
\expandafter\ifx\csname urlprefix\endcsname\relax\def\urlprefix{URL }\fi
\expandafter\ifx\csname href\endcsname\relax
  \def\href#1#2{#2} \def\path#1{#1}\fi

\bibitem{Parke:1986gb}
S.~J. Parke, T.~R. Taylor, {An Amplitude for $n$ Gluon Scattering}, Phys. Rev.
  Lett. 56 (1986) 2459.
\newblock \href {http://dx.doi.org/10.1103/PhysRevLett.56.2459}
  {\path{doi:10.1103/PhysRevLett.56.2459}}.

\bibitem{Berends:1987me}
F.~A. Berends, W.~T. Giele, {Recursive Calculations for Processes with n
  Gluons}, Nucl. Phys. B306 (1988) 759--808.
\newblock \href {http://dx.doi.org/10.1016/0550-3213(88)90442-7}
  {\path{doi:10.1016/0550-3213(88)90442-7}}.

\bibitem{Bern:1994zx}
Z.~Bern, L.~J. Dixon, D.~C. Dunbar, D.~A. Kosower, {One loop n point gauge
  theory amplitudes, unitarity and collinear limits}, Nucl. Phys. B 425 (1994)
  217--260.
\newblock \href {http://arxiv.org/abs/hep-ph/9403226}
  {\path{arXiv:hep-ph/9403226}}, \href
  {http://dx.doi.org/10.1016/0550-3213(94)90179-1}
  {\path{doi:10.1016/0550-3213(94)90179-1}}.

\bibitem{Bern:2011qt}
Z.~Bern, Y.-t. Huang, {Basics of Generalized Unitarity}, J. Phys. A 44 (2011)
  454003.
\newblock \href {http://arxiv.org/abs/1103.1869} {\path{arXiv:1103.1869}},
  \href {http://dx.doi.org/10.1088/1751-8113/44/45/454003}
  {\path{doi:10.1088/1751-8113/44/45/454003}}.

\bibitem{Witten:2003nn}
E.~Witten, \emph{Perturbative gauge theory as a string theory in twistor
  space}, Commun. Math. Phys. 252 (2004) 189--258.
\newblock \href {http://arxiv.org/abs/hep-th/0312171}
  {\path{arXiv:hep-th/0312171}}, \href
  {http://dx.doi.org/10.1007/s00220-004-1187-3}
  {\path{doi:10.1007/s00220-004-1187-3}}.

\bibitem{Britto:2004ap}
R.~Britto, F.~Cachazo, B.~Feng, {New recursion relations for tree amplitudes of
  gluons}, Nucl. Phys. B 715 (2005) 499--522.
\newblock \href {http://arxiv.org/abs/hep-th/0412308}
  {\path{arXiv:hep-th/0412308}}, \href
  {http://dx.doi.org/10.1016/j.nuclphysb.2005.02.030}
  {\path{doi:10.1016/j.nuclphysb.2005.02.030}}.

\bibitem{Britto:2005fq}
R.~Britto, F.~Cachazo, B.~Feng, E.~Witten, {Direct proof of tree-level
  recursion relation in Yang-Mills theory}, Phys.Rev.Lett. 94 (2005) 181602.
\newblock \href {http://arxiv.org/abs/hep-th/0501052}
  {\path{arXiv:hep-th/0501052}}, \href
  {http://dx.doi.org/10.1103/PhysRevLett.94.181602}
  {\path{doi:10.1103/PhysRevLett.94.181602}}.

\bibitem{Arkani-Hamed:2009ljj}
N.~Arkani-Hamed, F.~Cachazo, C.~Cheung, J.~Kaplan, {A Duality For The S
  Matrix}, JHEP 03 (2010) 020.
\newblock \href {http://arxiv.org/abs/0907.5418} {\path{arXiv:0907.5418}},
  \href {http://dx.doi.org/10.1007/JHEP03(2010)020}
  {\path{doi:10.1007/JHEP03(2010)020}}.

\bibitem{Mason:2009qx}
L.~J. Mason, D.~Skinner, {Dual Superconformal Invariance, Momentum Twistors and
  Grassmannians}, JHEP 11 (2009) 045.
\newblock \href {http://arxiv.org/abs/0909.0250} {\path{arXiv:0909.0250}},
  \href {http://dx.doi.org/10.1088/1126-6708/2009/11/045}
  {\path{doi:10.1088/1126-6708/2009/11/045}}.

\bibitem{Elvang:2013cua}
H.~Elvang, Y.-t. Huang, {Scattering Amplitudes}\href
  {http://arxiv.org/abs/1308.1697} {\path{arXiv:1308.1697}}.

\bibitem{Dixon:2013uaa}
L.~J. Dixon, {A brief introduction to modern amplitude methods}, in:
  {Theoretical Advanced Study Institute in Elementary Particle Physics}:
  {Particle Physics: The Higgs Boson and Beyond}, 2014, pp. 31--67.
\newblock \href {http://arxiv.org/abs/1310.5353} {\path{arXiv:1310.5353}},
  \href {http://dx.doi.org/10.5170/CERN-2014-008.31}
  {\path{doi:10.5170/CERN-2014-008.31}}.

\bibitem{Rosly:1996vr}
A.~A. Rosly, K.~G. Selivanov, {On amplitudes in selfdual sector of Yang-Mills
  theory}, Phys. Lett. B399 (1997) 135--140.
\newblock \href {http://arxiv.org/abs/hep-th/9611101}
  {\path{arXiv:hep-th/9611101}}, \href
  {http://dx.doi.org/10.1016/S0370-2693(97)00268-2}
  {\path{doi:10.1016/S0370-2693(97)00268-2}}.

\bibitem{Rosly:1997ap}
A.~A. Rosly, K.~G. Selivanov, {Gravitational SD perturbiner}\href
  {http://arxiv.org/abs/hep-th/9710196} {\path{arXiv:hep-th/9710196}}.

\bibitem{Selivanov:1997aq}
K.~G. Selivanov, {SD perturbiner in Yang-Mills + gravity}, Phys. Lett. B420
  (1998) 274--278.
\newblock \href {http://arxiv.org/abs/hep-th/9710197}
  {\path{arXiv:hep-th/9710197}}, \href
  {http://dx.doi.org/10.1016/S0370-2693(97)01514-1}
  {\path{doi:10.1016/S0370-2693(97)01514-1}}.

\bibitem{Selivanov:1997ts}
K.~G. Selivanov, {Gravitationally dressed Parke-Taylor amplitudes}, Mod. Phys.
  Lett. A12 (1997) 3087--3090.
\newblock \href {http://arxiv.org/abs/hep-th/9711111}
  {\path{arXiv:hep-th/9711111}}, \href
  {http://dx.doi.org/10.1142/S0217732397003204}
  {\path{doi:10.1142/S0217732397003204}}.

\bibitem{Bern:2008qj}
Z.~Bern, J.~J.~M. Carrasco, H.~Johansson, {New Relations for Gauge-Theory
  Amplitudes}, Phys. Rev. D78 (2008) 085011.
\newblock \href {http://arxiv.org/abs/0805.3993} {\path{arXiv:0805.3993}},
  \href {http://dx.doi.org/10.1103/PhysRevD.78.085011}
  {\path{doi:10.1103/PhysRevD.78.085011}}.

\bibitem{Mafra:2014oia}
C.~R. Mafra, O.~Schlotterer, {Multiparticle SYM equations of motion and pure
  spinor BRST blocks}, JHEP 07 (2014) 153.
\newblock \href {http://arxiv.org/abs/1404.4986} {\path{arXiv:1404.4986}},
  \href {http://dx.doi.org/10.1007/JHEP07(2014)153}
  {\path{doi:10.1007/JHEP07(2014)153}}.

\bibitem{Kawai:1985xq}
H.~Kawai, D.~C. Lewellen, S.~H.~H. Tye, {A Relation Between Tree Amplitudes of
  Closed and Open Strings}, Nucl. Phys. B269 (1986) 1--23.
\newblock \href {http://dx.doi.org/10.1016/0550-3213(86)90362-7}
  {\path{doi:10.1016/0550-3213(86)90362-7}}.

\bibitem{Bern:1998sv}
Z.~Bern, L.~J. Dixon, M.~Perelstein, J.~S. Rozowsky, {Multileg one loop gravity
  amplitudes from gauge theory}, Nucl. Phys. B 546 (1999) 423--479.
\newblock \href {http://arxiv.org/abs/hep-th/9811140}
  {\path{arXiv:hep-th/9811140}}, \href
  {http://dx.doi.org/10.1016/S0550-3213(99)00029-2}
  {\path{doi:10.1016/S0550-3213(99)00029-2}}.

\bibitem{Bjerrum-Bohr:2010pnr}
N.~E.~J. Bjerrum-Bohr, P.~H. Damgaard, T.~Sondergaard, P.~Vanhove, {The
  Momentum Kernel of Gauge and Gravity Theories}, JHEP 01 (2011) 001.
\newblock \href {http://arxiv.org/abs/1010.3933} {\path{arXiv:1010.3933}},
  \href {http://dx.doi.org/10.1007/JHEP01(2011)001}
  {\path{doi:10.1007/JHEP01(2011)001}}.

\bibitem{Mizera:2017cqs}
S.~Mizera, {Combinatorics and Topology of Kawai-Lewellen-Tye Relations}, JHEP
  08 (2017) 097.
\newblock \href {http://arxiv.org/abs/1706.08527} {\path{arXiv:1706.08527}},
  \href {http://dx.doi.org/10.1007/JHEP08(2017)097}
  {\path{doi:10.1007/JHEP08(2017)097}}.

\bibitem{Sondergaard:2011iv}
T.~Sondergaard, {Perturbative Gravity and Gauge Theory Relations: A Review},
  Adv. High Energy Phys. 2012 (2012) 726030.
\newblock \href {http://arxiv.org/abs/1106.0033} {\path{arXiv:1106.0033}},
  \href {http://dx.doi.org/10.1155/2012/726030}
  {\path{doi:10.1155/2012/726030}}.

\bibitem{Goldberger:2017frp}
W.~D. Goldberger, S.~G. Prabhu, J.~O. Thompson, {Classical gluon and graviton
  radiation from the bi-adjoint scalar double copy}, Phys. Rev. D 96~(6) (2017)
  065009.
\newblock \href {http://arxiv.org/abs/1705.09263} {\path{arXiv:1705.09263}},
  \href {http://dx.doi.org/10.1103/PhysRevD.96.065009}
  {\path{doi:10.1103/PhysRevD.96.065009}}.

\bibitem{Shen:2018ebu}
C.-H. Shen, {Gravitational Radiation from Color-Kinematics Duality}, JHEP 11
  (2018) 162.
\newblock \href {http://arxiv.org/abs/1806.07388} {\path{arXiv:1806.07388}},
  \href {http://dx.doi.org/10.1007/JHEP11(2018)162}
  {\path{doi:10.1007/JHEP11(2018)162}}.

\bibitem{Cheung:2018wkq}
C.~Cheung, I.~Z. Rothstein, M.~P. Solon, {From Scattering Amplitudes to
  Classical Potentials in the Post-Minkowskian Expansion}, Phys. Rev. Lett.
  121~(25) (2018) 251101.
\newblock \href {http://arxiv.org/abs/1808.02489} {\path{arXiv:1808.02489}},
  \href {http://dx.doi.org/10.1103/PhysRevLett.121.251101}
  {\path{doi:10.1103/PhysRevLett.121.251101}}.

\bibitem{LIGOScientific:2016aoc}
B.~P. Abbott, et~al., {Observation of Gravitational Waves from a Binary Black
  Hole Merger}, Phys. Rev. Lett. 116~(6) (2016) 061102.
\newblock \href {http://arxiv.org/abs/1602.03837} {\path{arXiv:1602.03837}},
  \href {http://dx.doi.org/10.1103/PhysRevLett.116.061102}
  {\path{doi:10.1103/PhysRevLett.116.061102}}.

\bibitem{LIGOScientific:2017vwq}
B.~P. Abbott, et~al., {GW170817: Observation of Gravitational Waves from a
  Binary Neutron Star Inspiral}, Phys. Rev. Lett. 119~(16) (2017) 161101.
\newblock \href {http://arxiv.org/abs/1710.05832} {\path{arXiv:1710.05832}},
  \href {http://dx.doi.org/10.1103/PhysRevLett.119.161101}
  {\path{doi:10.1103/PhysRevLett.119.161101}}.

\bibitem{Shi:2021qsb}
C.~Shi, J.~Plefka, {Classical Double Copy of Worldline Quantum Field
  Theory}\href {http://arxiv.org/abs/2109.10345} {\path{arXiv:2109.10345}}.

\bibitem{Bjerrum-Bohr:2009ulz}
N.~E.~J. Bjerrum-Bohr, P.~H. Damgaard, P.~Vanhove, {Minimal Basis for Gauge
  Theory Amplitudes}, Phys. Rev. Lett. 103 (2009) 161602.
\newblock \href {http://arxiv.org/abs/0907.1425} {\path{arXiv:0907.1425}},
  \href {http://dx.doi.org/10.1103/PhysRevLett.103.161602}
  {\path{doi:10.1103/PhysRevLett.103.161602}}.

\bibitem{Stieberger:2009hq}
S.~Stieberger, {Open \& Closed vs. Pure Open String Disk Amplitudes}\href
  {http://arxiv.org/abs/0907.2211} {\path{arXiv:0907.2211}}.

\bibitem{Bjerrum-Bohr:2010mia}
N.~E.~J. Bjerrum-Bohr, P.~H. Damgaard, T.~Sondergaard, P.~Vanhove, {Monodromy
  and Jacobi-like Relations for Color-Ordered Amplitudes}, JHEP 06 (2010) 003.
\newblock \href {http://arxiv.org/abs/1003.2403} {\path{arXiv:1003.2403}},
  \href {http://dx.doi.org/10.1007/JHEP06(2010)003}
  {\path{doi:10.1007/JHEP06(2010)003}}.

\bibitem{Bern:2019prr}
Z.~Bern, J.~J. Carrasco, M.~Chiodaroli, H.~Johansson, R.~Roiban, {The Duality
  Between Color and Kinematics and its Applications}\href
  {http://arxiv.org/abs/1909.01358} {\path{arXiv:1909.01358}}.

\bibitem{Bern:2010ue}
Z.~Bern, J.~J.~M. Carrasco, H.~Johansson, {Perturbative Quantum Gravity as a
  Double Copy of Gauge Theory}, Phys. Rev. Lett. 105 (2010) 061602.
\newblock \href {http://arxiv.org/abs/1004.0476} {\path{arXiv:1004.0476}},
  \href {http://dx.doi.org/10.1103/PhysRevLett.105.061602}
  {\path{doi:10.1103/PhysRevLett.105.061602}}.

\bibitem{Carrasco:2011mn}
J.~J.~M. Carrasco, H.~Johansson, {Five-Point Amplitudes in N=4 Super-Yang-Mills
  Theory and N=8 Supergravity}, Phys. Rev. D 85 (2012) 025006.
\newblock \href {http://arxiv.org/abs/1106.4711} {\path{arXiv:1106.4711}},
  \href {http://dx.doi.org/10.1103/PhysRevD.85.025006}
  {\path{doi:10.1103/PhysRevD.85.025006}}.

\bibitem{Carrasco:2011hw}
J.~J.~M. Carrasco, H.~Johansson, {Generic multiloop methods and application to
  N=4 super-Yang-Mills}, J. Phys. A 44 (2011) 454004.
\newblock \href {http://arxiv.org/abs/1103.3298} {\path{arXiv:1103.3298}},
  \href {http://dx.doi.org/10.1088/1751-8113/44/45/454004}
  {\path{doi:10.1088/1751-8113/44/45/454004}}.

\bibitem{Bern:2011rj}
Z.~Bern, C.~Boucher-Veronneau, H.~Johansson, {N \ensuremath{>}= 4 Supergravity
  Amplitudes from Gauge Theory at One Loop}, Phys. Rev. D 84 (2011) 105035.
\newblock \href {http://arxiv.org/abs/1107.1935} {\path{arXiv:1107.1935}},
  \href {http://dx.doi.org/10.1103/PhysRevD.84.105035}
  {\path{doi:10.1103/PhysRevD.84.105035}}.

\bibitem{Boucher-Veronneau:2011rlc}
C.~Boucher-Veronneau, L.~J. Dixon, {N \ensuremath{>}- 4 Supergravity Amplitudes
  from Gauge Theory at Two Loops}, JHEP 12 (2011) 046.
\newblock \href {http://arxiv.org/abs/1110.1132} {\path{arXiv:1110.1132}},
  \href {http://dx.doi.org/10.1007/JHEP12(2011)046}
  {\path{doi:10.1007/JHEP12(2011)046}}.

\bibitem{Naculich:2011my}
S.~G. Naculich, H.~Nastase, H.~J. Schnitzer, {Linear relations between N
  \ensuremath{>}= 4 supergravity and subleading-color SYM amplitudes}, JHEP 01
  (2012) 041.
\newblock \href {http://arxiv.org/abs/1111.1675} {\path{arXiv:1111.1675}},
  \href {http://dx.doi.org/10.1007/JHEP01(2012)041}
  {\path{doi:10.1007/JHEP01(2012)041}}.

\bibitem{Bern:2012uf}
Z.~Bern, J.~J.~M. Carrasco, L.~J. Dixon, H.~Johansson, R.~Roiban, {Simplifying
  Multiloop Integrands and Ultraviolet Divergences of Gauge Theory and Gravity
  Amplitudes}, Phys. Rev. D 85 (2012) 105014.
\newblock \href {http://arxiv.org/abs/1201.5366} {\path{arXiv:1201.5366}},
  \href {http://dx.doi.org/10.1103/PhysRevD.85.105014}
  {\path{doi:10.1103/PhysRevD.85.105014}}.

\bibitem{Bargheer:2012gv}
T.~Bargheer, S.~He, T.~McLoughlin, {New Relations for Three-Dimensional
  Supersymmetric Scattering Amplitudes}, Phys. Rev. Lett. 108 (2012) 231601.
\newblock \href {http://arxiv.org/abs/1203.0562} {\path{arXiv:1203.0562}},
  \href {http://dx.doi.org/10.1103/PhysRevLett.108.231601}
  {\path{doi:10.1103/PhysRevLett.108.231601}}.

\bibitem{Burger:2021wss}
D.~J. Burger, W.~T. Emond, N.~Moynihan, {Anyons and the Double Copy}\href
  {http://arxiv.org/abs/2103.10416} {\path{arXiv:2103.10416}}.

\bibitem{Bern1991-1669}
Z.~Bern, D.~A. Kosower,
  \href{https://link.aps.org/doi/10.1103/PhysRevLett.66.1669}{{Efficient
  calculation of one-loop QCD amplitudes}}, Phys. Rev. Lett. 66 (1991)
  1669--1672.
\newblock \href {http://dx.doi.org/10.1103/PhysRevLett.66.1669}
  {\path{doi:10.1103/PhysRevLett.66.1669}}.
\newline\urlprefix\url{https://link.aps.org/doi/10.1103/PhysRevLett.66.1669}

\bibitem{bern1991color}
Z.~Bern, D.~A. Kosower, {Color decomposition of one loop amplitudes in gauge
  theories}, Nucl. Phys. B 362 (1991) 389--448.
\newblock \href {http://dx.doi.org/10.1016/0550-3213(91)90567-H}
  {\path{doi:10.1016/0550-3213(91)90567-H}}.

\bibitem{bern1992computation}
Z.~Bern, D.~A. Kosower, {The Computation of loop amplitudes in gauge theories},
  Nucl. Phys. B 379 (1992) 451--561.
\newblock \href {http://dx.doi.org/10.1016/0550-3213(92)90134-W}
  {\path{doi:10.1016/0550-3213(92)90134-W}}.

\bibitem{Strassler:1992zr}
M.~J. Strassler, {Field theory without Feynman diagrams: One loop effective
  actions}, Nucl. Phys. B 385 (1992) 145--184.
\newblock \href {http://arxiv.org/abs/hep-ph/9205205}
  {\path{arXiv:hep-ph/9205205}}, \href
  {http://dx.doi.org/10.1016/0550-3213(92)90098-V}
  {\path{doi:10.1016/0550-3213(92)90098-V}}.

\bibitem{Strassler:1992nc}
M.~J. Strassler, {Field theory without Feynman diagrams: A Demonstration using
  actions induced by heavy particles}, Unpublished.

\bibitem{Bern:1993wt}
Z.~Bern, D.~C. Dunbar, T.~Shimada, {String based methods in perturbative
  gravity}, Phys. Lett. B 312 (1993) 277--284.
\newblock \href {http://arxiv.org/abs/hep-th/9307001}
  {\path{arXiv:hep-th/9307001}}, \href
  {http://dx.doi.org/10.1016/0370-2693(93)91081-W}
  {\path{doi:10.1016/0370-2693(93)91081-W}}.

\bibitem{Dunbar:1994bn}
D.~C. Dunbar, P.~S. Norridge, {Calculation of graviton scattering amplitudes
  using string based methods}, Nucl. Phys. B 433 (1995) 181--208.
\newblock \href {http://arxiv.org/abs/hep-th/9408014}
  {\path{arXiv:hep-th/9408014}}, \href
  {http://dx.doi.org/10.1016/0550-3213(94)00385-R}
  {\path{doi:10.1016/0550-3213(94)00385-R}}.

\bibitem{Ahmadiniaz:2021fey}
N.~Ahmadiniaz, F.~M. Balli, C.~Lopez-Arcos, A.~Quintero~Velez, C.~Schubert,
  {Color-kinematics duality from the Bern-Kosower formalism}, Phys. Rev. D
  104~(4) (2021) L041702.
\newblock \href {http://arxiv.org/abs/2105.06745} {\path{arXiv:2105.06745}},
  \href {http://dx.doi.org/10.1103/PhysRevD.104.L041702}
  {\path{doi:10.1103/PhysRevD.104.L041702}}.

\bibitem{Mizera:2018jbh}
S.~Mizera, B.~Skrzypek, {Perturbiner Methods for Effective Field Theories and
  the Double Copy}, JHEP 10 (2018) 018.
\newblock \href {http://arxiv.org/abs/1809.02096} {\path{arXiv:1809.02096}},
  \href {http://dx.doi.org/10.1007/JHEP10(2018)018}
  {\path{doi:10.1007/JHEP10(2018)018}}.

\bibitem{Cho:2021nim}
K.~Cho, K.~Kim, K.~Lee, {The Off-Shell Recursion for Gravity and the Classical
  Double Copy for currents}\href {http://arxiv.org/abs/2109.06392}
  {\path{arXiv:2109.06392}}.

\bibitem{Schubert:1997ph}
C.~Schubert, {The Structure of the Bern-Kosower integrand for the N gluon
  amplitude}, Eur. Phys. J. C 5 (1998) 693--699.
\newblock \href {http://arxiv.org/abs/hep-th/9710067}
  {\path{arXiv:hep-th/9710067}}, \href
  {http://dx.doi.org/10.1007/s100520050311} {\path{doi:10.1007/s100520050311}}.

\bibitem{Ahmadiniaz:2012ie}
N.~Ahmadiniaz, C.~Schubert, V.~M. Villanueva, {String-inspired representations
  of photon/gluon amplitudes}, JHEP 01 (2013) 132.
\newblock \href {http://arxiv.org/abs/1211.1821} {\path{arXiv:1211.1821}},
  \href {http://dx.doi.org/10.1007/JHEP01(2013)132}
  {\path{doi:10.1007/JHEP01(2013)132}}.

\bibitem{Schubert2001-73}
C.~Schubert,
  \href{http://www.sciencedirect.com/science/article/pii/S0370157301000138}{Perturbative
  quantum field theory in the string-inspired formalism}, Physics Reports
  355~(2) (2001) 73 -- 234.
\newblock \href {http://arxiv.org/abs/hep-th/0101036}
  {\path{arXiv:hep-th/0101036}}, \href
  {http://dx.doi.org/https://doi.org/10.1016/S0370-1573(01)00013-8}
  {\path{doi:https://doi.org/10.1016/S0370-1573(01)00013-8}}.
\newline\urlprefix\url{http://www.sciencedirect.com/science/article/pii/S0370157301000138}

\bibitem{Bridges:2019siz}
E.~Bridges, C.~R. Mafra, {Algorithmic construction of SYM multiparticle
  superfields in the BCJ gauge}, JHEP 10 (2019) 022.
\newblock \href {http://arxiv.org/abs/1906.12252} {\path{arXiv:1906.12252}},
  \href {http://dx.doi.org/10.1007/JHEP10(2019)022}
  {\path{doi:10.1007/JHEP10(2019)022}}.

\bibitem{Kleiss:1988ne}
R.~Kleiss, H.~Kuijf, {Multi - Gluon Cross-sections and Five Jet Production at
  Hadron Colliders}, Nucl. Phys. B312 (1989) 616--644.
\newblock \href {http://dx.doi.org/10.1016/0550-3213(89)90574-9}
  {\path{doi:10.1016/0550-3213(89)90574-9}}.

\bibitem{LAGQV21}
C.~Lopez-Arcos, V.~Guarin~Escudero, A.~Quintero~Velez, Double copy and
  multiparticle solutions to the non-abelian {N}avier-{S}tokes equations, to
  appear (2021).

\bibitem{Wu:2021exa}
K.~Wu, Y.-J. Du, {Off-shell extended graphic rule and the expansion of
  Berends-Giele currents in Yang-Mills theory}\href
  {http://arxiv.org/abs/2109.14462} {\path{arXiv:2109.14462}}.

\bibitem{Ben-Shahar:2021doh}
M.~Ben-Shahar, M.~Guillen, {10D Super-Yang-Mills Scattering Amplitudes From Its
  Pure Spinor Action}\href {http://arxiv.org/abs/2108.11708}
  {\path{arXiv:2108.11708}}.

\bibitem{Bjerrum-Bohr:2003hzh}
N.~E.~J. Bjerrum-Bohr, {String theory and the mapping of gravity into gauge
  theory}, Phys. Lett. B 560 (2003) 98--107.
\newblock \href {http://arxiv.org/abs/hep-th/0302131}
  {\path{arXiv:hep-th/0302131}}, \href
  {http://dx.doi.org/10.1016/S0370-2693(03)00373-3}
  {\path{doi:10.1016/S0370-2693(03)00373-3}}.

\bibitem{Metsaev:1986yb}
R.~R. Metsaev, A.~A. Tseytlin, {Curvature Cubed Terms in String Theory
  Effective Actions}, Phys. Lett. B 185 (1987) 52--58.
\newblock \href {http://dx.doi.org/10.1016/0370-2693(87)91527-9}
  {\path{doi:10.1016/0370-2693(87)91527-9}}.

\bibitem{Broedel:2012rc}
J.~Broedel, L.~J. Dixon, {Color-kinematics duality and double-copy construction
  for amplitudes from higher-dimension operators}, JHEP 10 (2012) 091.
\newblock \href {http://arxiv.org/abs/1208.0876} {\path{arXiv:1208.0876}},
  \href {http://dx.doi.org/10.1007/JHEP10(2012)091}
  {\path{doi:10.1007/JHEP10(2012)091}}.

\bibitem{Garozzo:2018uzj}
L.~M. Garozzo, L.~Queimada, O.~Schlotterer, {Berends-Giele currents in
  Bern-Carrasco-Johansson gauge for $F^3$- and $F^4$-deformed Yang-Mills
  amplitudes}, JHEP 02 (2019) 078.
\newblock \href {http://arxiv.org/abs/1809.08103} {\path{arXiv:1809.08103}},
  \href {http://dx.doi.org/10.1007/JHEP02(2019)078}
  {\path{doi:10.1007/JHEP02(2019)078}}.

\bibitem{Mafra:2016ltu}
C.~R. Mafra, {Berends-Giele recursion for double-color-ordered amplitudes},
  JHEP 07 (2016) 080.
\newblock \href {http://arxiv.org/abs/1603.09731} {\path{arXiv:1603.09731}},
  \href {http://dx.doi.org/10.1007/JHEP07(2016)080}
  {\path{doi:10.1007/JHEP07(2016)080}}.

\bibitem{Lopez-Arcos:2019hvg}
C.~Lopez-Arcos, A.~Quintero~V\'elez, {$L_{\infty}$-algebras and the perturbiner
  expansion}, JHEP 11 (2019) 010.
\newblock \href {http://arxiv.org/abs/1907.12154} {\path{arXiv:1907.12154}},
  \href {http://dx.doi.org/10.1007/JHEP11(2019)010}
  {\path{doi:10.1007/JHEP11(2019)010}}.

\bibitem{Lee:2015upy}
S.~Lee, C.~R. Mafra, O.~Schlotterer, {Non-linear gauge transformations in
  $D=10$ SYM theory and the BCJ duality}, JHEP 03 (2016) 090.
\newblock \href {http://arxiv.org/abs/1510.08843} {\path{arXiv:1510.08843}},
  \href {http://dx.doi.org/10.1007/JHEP03(2016)090}
  {\path{doi:10.1007/JHEP03(2016)090}}.

\bibitem{Cachazo:2013iaa}
F.~Cachazo, S.~He, E.~Y. Yuan, \emph{Scattering in Three Dimensions from
  Rational Maps}, JHEP 10 (2013) 141.
\newblock \href {http://arxiv.org/abs/1306.2962} {\path{arXiv:1306.2962}},
  \href {http://dx.doi.org/10.1007/JHEP10(2013)141}
  {\path{doi:10.1007/JHEP10(2013)141}}.

\bibitem{White:2020sfn}
C.~D. White, {Twistorial Foundation for the Classical Double Copy}, Phys. Rev.
  Lett. 126~(6) (2021) 061602.
\newblock \href {http://arxiv.org/abs/2012.02479} {\path{arXiv:2012.02479}},
  \href {http://dx.doi.org/10.1103/PhysRevLett.126.061602}
  {\path{doi:10.1103/PhysRevLett.126.061602}}.

\bibitem{Gomez:2020vat}
H.~Gomez, R.~L. Jusinskas, C.~Lopez-Arcos, A.~Quintero~V\'{e}lez, {The
  $L_{\infty}$ structure of gauge theories with matter}, JHEP 02 (2021) 093.
\newblock \href {http://arxiv.org/abs/2011.09528} {\path{arXiv:2011.09528}},
  \href {http://dx.doi.org/10.1007/JHEP02(2021)093}
  {\path{doi:10.1007/JHEP02(2021)093}}.

\bibitem{Johansson:2019dnu}
H.~Johansson, A.~Ochirov, {Double copy for massive quantum particles with
  spin}, JHEP 09 (2019) 040.
\newblock \href {http://arxiv.org/abs/1906.12292} {\path{arXiv:1906.12292}},
  \href {http://dx.doi.org/10.1007/JHEP09(2019)040}
  {\path{doi:10.1007/JHEP09(2019)040}}.

\bibitem{Gomez:2021shh}
H.~Gomez, R.~L. Jusinskas, {Multiparticle solutions to Einstein's
  equations}\href {http://arxiv.org/abs/2106.12584} {\path{arXiv:2106.12584}}.

\end{thebibliography}
\end{document}